\documentclass[11pt,a4paper]{article}

\usepackage{amsmath}
\usepackage{mathtools}
\usepackage{nccmath}
\usepackage{amsthm}
\usepackage{amssymb}
\usepackage{amsfonts}
\usepackage{anyfontsize}
\usepackage{mathrsfs}
\usepackage{bbm}
\usepackage{graphicx}
\usepackage{caption}
\usepackage{bookmark,hyperref}
\usepackage{12many}
\usepackage{parskip}
\usepackage{setspace}
\usepackage{multirow}
\usepackage{arydshln}
\usepackage{anyfontsize}
\usepackage{changepage}
\usepackage{todonotes}
\usepackage{braket}
\usepackage{dirtytalk}
\usepackage{MnSymbol}

\numberwithin{equation}{section}
\usepackage{fullpage}

\begin{document}
\begin{titlepage}
 \thispagestyle{empty}
 \begin{flushright}
 \hfill{Imperial-TP-2024-CH-02 }\\
   %  \hfill{ITP-UU-14/29 }\\
      % \hfill{CPHT-RR098.1214 }\\
 \end{flushright}

 \vspace{30pt}

 \begin{center}
     
  {\fontsize{20}{24} \bf {Exact moduli spaces for $\mathcal{N}=2, D=5$ \\ \vspace{0.7cm} freely acting orbifolds}}

     \vspace{30pt}
{\fontsize{13}{16}\selectfont {George Gkountoumis$^1$, Chris Hull$^2$ and Stefan Vandoren$^1$}} \\[10mm]

{\small\it
${}^1$ Institute for Theoretical Physics {and} Center for Extreme Matter and Emergent Phenomena \\
Utrecht University, 3508 TD Utrecht, The Netherlands \\[3mm]

${}^2$ %Imperial College, London, UK \\[3mm]
{{The Blackett Laboratory}},
{{Imperial College London}}\\
{{Prince Consort Road}}, 
{{London SW7 2AZ, U.K.}}\\[3mm]}

\vspace{1.5cm}

{\bf Abstract}

\vspace{0.3cm}
   
\begin{adjustwidth}{12pt}{12pt}

We use freely acting asymmetric orbifolds
of type IIB string theory to construct a class of  theories in five dimensions with eight supercharges whose moduli spaces for vector multiplets and hypermultiplets can be determined exactly. We argue that no quantum corrections  to these moduli spaces arise.  We focus   on examples in which
 all  moduli are in the NS-NS sector, while all fields from the R-R sector become massive. The full symmetry group of the moduli space is then determined  by the subgroup of the
 T-duality group that survives  the orbifold action. We illustrate this for  freely acting orbifolds of type IIB string theory on $T^5$ with $0,1$ or $2$ hypermultiplets. 

\end{adjustwidth}

\end{center}

\vspace{20pt}

\newcommand\blfootnote[1]{%
  \begingroup
  \renewcommand\thefootnote{}\footnote{#1}%
  \addtocounter{footnote}{-1}%
  \endgroup}

\blfootnote{g.gkountoumis@uu.nl \quad c.hull@imperial.ac.uk \quad s.j.g.vandoren@uu.nl}

\noindent

\end{titlepage}

\begin{spacing}{1.15}
\tableofcontents
\end{spacing}

\section{Introduction}

The compactification of M-theory  on a Calabi-Yau threefold (CY$_3$)   gives a five dimensional theory with eight supercharges, that is  with $\mathcal{N}=2$ supersymmetry in $D=5$. The spectrum contains $h_{1,2}+1$ massless hypermultiplets and $h_{1,1}-1$ vector multiplets.  Generically, the hypermultiplet moduli space receives quantum corrections coming from M2- and M5-brane instantons \cite{Becker:1995kb}. These deform the metric on the classical moduli space to the full quantum corrected metric on the hypermultiplet moduli space. The vector multiplet moduli space is determined by the intersection numbers of the CY$_3$ \cite{Cadavid:1995bk}. It was described in \cite{witten:1996qb} and exhibits a rich structure of topology-changing transitions.

After further reduction on a circle $S^1$ to 4 dimensions, the theory is dual to type IIA string theory compactified on the same CY$_3$. The numbers of four-dimensional hypermultiplets  and vector multiplets  are now given by $n_H=h_{1,2}+1$ and $n_V=h_{1,1}$ respectively and the dilaton sits in a hypermultiplet. There are quantum corrections depending on the string coupling constant $\lambda _s$ that deform the  hypermultiplet moduli space metric, both perturbatively and non-perturbatively.
The vector multiplet moduli space metric receives world-sheet instanton corrections in the IIA theory. In the M-theory picture, these arise from membranes wrapping holomorphic curves in the CY$_3$ times $S^1$. Similarly, type IIB string theory compactifications on a CY$_3$ yield $n_V=h_{1,2}$ vector multiplets and $n_H=h_{1,1}+1$ hypermultiplets in $D=4$. As in type IIA, the dilaton sits in a hypermultiplet. For some literature on quantum corrections to hypermultiplet moduli spaces in type II, see e.g.\ \cite{Antoniadis:2003sw,Robles-Llana:2006vct} for  loop corrections, and \cite{Robles-Llana:2006hby,Robles-Llana:2007bbv,Alexandrov:2008gh,Alexandrov:2021shf,Alexandrov:2021dyl,Alexandrov:2023hiv} for a partial list of references on instanton corrections.

In this paper, we consider $\mathcal{N}=2, D=5$ string models based on freely acting asymmetric orbifold compactifications of type IIB on $T^5$  \cite{Gkountoumis:2023fym}. They constitute a special corner in the string landscape of non-geometric compactifications in which supersymmetry is broken spontaneously, in the case at hand from $\mathcal{N}=8$ to $\mathcal{N}=2$, which is the minimal amount of supersymmetry in five dimensions. The fact that $\mathcal{N}=8$ supersymmetry is spontaneously broken instead of being explicitly broken
results in special properties arising from the hidden 
$\mathcal{N}=8$ supersymmetry.

We will determine the moduli spaces exactly and argue that they receive no quantum corrections. The dilaton sits in a vector multiplet, just as it does for the heterotic string theory on $K3\times S^1$ in five dimensions\footnote{The heterotic string on $K3\times S^1$ is dual to M-theory on a CY$_3$ \cite{Cadavid:1995bk,Papadopoulos:1995da,Antoniadis:1995vz,Ferrara:1996hh} and has the dilaton in a vector multiplet. The heterotic hypermultiplet moduli space is classically exact, and sums up all non-perturbative corrections from M-theory. The heterotic models contain many more hypermultiplets than our models and furthermore receive loop corrections to the vector multiplet moduli space, see e.g.\ \cite{Antoniadis:1995vz,Bonetti:2013cza}. A key difference with our models is that the heterotic string on $K3\times S^1$ is not obtained from a spontaneously broken $\mathcal{N}=8$ theory.}. Our strategy in this work is to first determine the classical hypermultiplet and vector multiplet moduli spaces by finding the duality group of the orbifold, that is the subgroup of the U-duality group of $T^5$ that commutes with the orbifold action. 
Then we construct further freely acting orbifolds that are the strong coupling duals of the  orbifolds we are considering.
Then, we argue that no quantum corrections to the moduli space metrics can arise. Quantum corrections to the moduli space metrics would result in  modifications to the 
\emph{classical} moduli space metrics of the dual theories and these are ruled out as we determine these classical metrics explicitly.
This absence of quantum corrections might also be thought of 
as a consequence of the hidden $\mathcal{N}=8$ and the dualities that remain after orbifolding. 

The models we consider here have only a few moduli, and the most extreme case has no hypermultiplet scalars and only two scalars from the vector multiplets. All this makes these models quite different from non-freely acting orbifolds and the geometric compactifications discussed above.

Models with spontaneous breaking from $\mathcal{N}=4$ to $\mathcal{N}=2$ supersymmetry based on freely acting orbifolds have been considered in e.g.\ \cite{Vafa:1995gm,Kiritsis:1997ca} and have been used to construct dual heterotic/type II pairs. Models with spontaneous breaking from $\mathcal{N}=8$ to $\mathcal{N}=4$ have been considered in e.g.\
\cite{sen1995dual,Gregori:1997hi}. Most of this literature deals with four dimensions, whereas our interest lies in five. One  of our motivations is that type II string theories with $\mathcal{N}=2$ supersymmetry in $D=5$ have not been explored to any great extent, partly because there seems to be no straightforward geometric   compactifications with $\mathcal{N}=2$ Minkowski vacua in five dimensions. Models with spontaneous breaking from $\mathcal{N}=8$ to $\mathcal{N}=2$ in $D=5$ from asymmetric orbifolds have been recently discussed in \cite{Gkountoumis:2023fym,Baykara:2023plc}, but no detailed analysis of the moduli spaces was considered. In this paper, we fill this gap. 

The orbifold examples we consider here  have the special property that all R-R fields are  massive, and therefore the moduli space consists purely of NS-NS fields. This is mostly a convenient rather than an essential ingredient to determine the moduli spaces. As we will show explicitly, by acting with U-duality transformations on our models, which have only NS-NS massless fields, we can produce models with massless R-R fields as well. One consequence of having only massless NS-NS fields is that there are no R-R charges and so no BPS D-branes carrying R-R  charges. In addition, the determination of the moduli space only involves calculations within the T-duality group instead of the full U-duality group, which makes the computations easier. 

In section \ref{FAO}, we review and discuss general aspects of freely acting asymmetric orbifolds of type IIB on $T^5$. In sections \ref{OSS} and \ref{EFFSUGRA}, we discuss the structure of the U-duality and R-symmetry groups surviving after the orbifold, as well as the effective supergravity theory, and give arguments for the absence of quantum corrections to the vector multiplet moduli space based on the fact that our theories have spontaneously broken $\mathcal{N}=8$ supersymmetry. Then, in section \ref{N=2 models in 5d with exact moduli spaces}, we construct various examples of $\mathcal{N}=2, D=5$ models and compute the classical hypermultiplet and vector multiplet moduli spaces. The examples we consider involve models with only a few hypermultiplets, $n_H=0,1$ or $2$, but can be generalised to higher $n_H$. The case with no hypermultiplets ($n_H=0$) was recently also discussed in \cite{Baykara:2023plc}, which has inspired the work presented here. In section \ref{dual pairs}, we discuss dual pairs based on the work of Sen and Vafa \cite{sen1995dual}, which give information about the vector multiplet moduli space in the strong string coupling regime, and also provide an understanding of the moduli space of some dual models that do contain massless R-R fields. Then, in section \ref{absence} we use this information to argue that there are no quantum corrections to the vector multiplet moduli spaces in all of our models, supporting our expectations from spontaneously broken $\mathcal{N}=8$ supersymmetry. We end with a discussion and outlook in section \ref{conclusions}.

\section{Freely acting orbifolds}\label{FAO}

In this section we briefly discuss freely acting asymmetric orbifold compactifications of type IIB string theory. This section is mostly based on \cite{Gkountoumis:2023fym}.

\subsection{General remarks}
The orbifolds we are interested 
in here have target spaces of the form 
\begin{equation}\label{background}
\mathbb{R}^{1,4}\times  S^1\times T^4\ , \end{equation}
identified under the action of a $\mathbb{Z}_p$ symmetry.
This requires being at a point in the moduli space in which the fields on $T^4$ are invariant under the action of 
$\mathbb{Z}_p$. 
We then orbifold by a $\mathbb{Z}_p$ given by this $T^4$ symmetry combined with a shift by $2\pi \mathcal{R}/p$ on the circle $S^1$ with radius $\mathcal{R}$ which makes the orbifolds freely acting. Freely acting orbifolds have no fixed points and, at generic points in the moduli space, all states coming from the twisted sectors are massive. Furthermore, in contrast to non-freely acting orbifolds, supersymmetry is spontaneously broken instead of being explicitly broken, manifested by the fact that gravitini  become massive instead of being projected out. 

Symmetric orbifolds arise when the $\mathbb{Z}_p$ action on $T^4$ is a geometric discrete symmetry of $T^4$, generated by a diffeomorphism on $T^4$, i.e.\ by an element in GL(4;$\mathbb{Z}$). As shown in \cite{Gkountoumis:2023fym}, they only lead to $\mathcal{N}=4$ or $\mathcal{N}=0$ supersymmetry in 5 dimensions. For asymmetric orbifolds, the $\mathbb{Z}_p$ group  acts as a T-duality transformation on the $T^4$ CFT. The T-duality group for superstrings on $T^4$ is Spin$(4,4;\mathbb{Z})$, a discrete subgroup
of the double cover Spin(4,4) of $\text{SO}(4,4)$, as the D-brane states transform as a spinor representation of Spin$(4,4)$ \cite{Hull:1994ys}. Asymmetric orbifolds can lead to $\mathcal{N}=6,4,2$ or $0$ supersymmetry.

For the purposes of this work we will consider freely acting orbifolds that preserve 8 supercharges in $5D$, that is $\mathcal{N}=2$ supersymmetry.  In the class of  models that were studied in \cite{Gkountoumis:2023fym}, these are necessarily asymmetric orbifolds. A general property of these asymmetric orbifolds is that the dilaton always sits in a vector multiplet, as it is a singlet under the
$\text{SU}(2)_R$ $R$-symmetry in $\mathcal{N}=2, D=5$  (see e.g.\ \cite{Gregori:1999ny,Gregori:1999ns}). In fact, some of our models have no hypermultiplets, so the dilaton, which always remains massless in our constructions, should necessarily sit in a vector multiplet.

\subsection{Asymmetric orbifolds}

For the construction of asymmetric orbifolds we follow the procedure presented in the original papers \cite{Narain:1986qm,narain1991asymmetric}. In general, upon compactification on $T^5$ the momentum and winding numbers take values in a Narain lattice $\Gamma^{5,5}$, which is an even, self-dual lattice \cite{narain1989new}. For our purposes we decompose $T^5=T^4\times S^1$ and correspondingly decompose the lattice as $\Gamma^{4,4}(\mathcal{G})\oplus \Gamma^{1,1}$, where $\Gamma^{4,4}(\mathcal{G})$ is an even, self-dual Lie algebra lattice admitting purely left and right-moving  symmetries. Such a lattice can be realised at special points in the moduli space as 
\begin{equation}
    \Gamma^{4,4}(\mathcal{G}) \equiv \{(p_L,p_R)|\,p_L \in \Lambda_W(\mathcal{G}),\, p_R \in \Lambda_W(\mathcal{G}),\, p_L-p_R \in \Lambda_R(\mathcal{G})\}\,.
\end{equation}
Here $\mathcal{G}$ is a Lie algebra of rank four and $\Lambda_W(\mathcal{G})$, $\Lambda_R(\mathcal{G})$ denote the weight and root lattices of $\mathcal{G}$, respectively. Now, the orbifold acts as a rotation on $\Gamma^{4,4}(\mathcal{G})$ and as a shift on $\Gamma^{1,1}$. Here we will only consider rotations $\mathcal{M}_{\theta}=(\mathcal{N}_L,\mathcal{N}_R) \in \text{SO}(4)_L\times \text{SO}(4)_R \subset \text{SO}(4,4)\,$ that do not mix left and right-movers. 
For a $\mathbb{Z}_p$ orbifold, we require that the rotation satisfies 
$(\mathcal{M}_{\theta})^p=1$.
Consistency of the asymmetric orbifold requires that the rotation is in the automorphism group of the lattice\footnote{A discussion on consistent rotations can be found e.g.\ in \cite{lerchie1989lattices}, cf. appendix B.}. Since the orbifold acts as a shift on $\Gamma^{1,1}$, it leaves $\Gamma^{1,1}$ invariant\footnote{Due to the shift, momentum states pick up a phase in the untwisted sector. In the twisted sectors states become massive.}. On the other hand, $\Gamma^{4,4}(\mathcal{G})$ is not in general invariant under rotations $\mathcal{M}_{\theta}$. If there exists a sublattice $I\subset \Gamma^{4,4}(\mathcal{G})$ that is invariant under the orbifold action, it is given by
\begin{equation}
    I \equiv \{ p \in \Gamma^{4,4}(\mathcal{G})\,|\, \mathcal{M}_{\theta}\cdot p = p\}\,.
\end{equation}
Then the complete  sublattice that is invariant under the 
orbifold action
is 
\begin{equation}
    \hat{I}=I\oplus \Gamma^{1,1}\,.
\end{equation}

It will be useful to determine the orbifold action on the world-sheet fields. We denote the $S^1$ coordinate by $Z$ and the  four real $T^4$ coordinates by $Y^m, m=1,\ldots 4$, which we combine into two complex coordinates
 $W^i = \tfrac{1}{\sqrt{2}}(Y^{2i-1}+iY^{2i})$ with $i=1,2$. We parametrize the rotations $(\mathcal{N}_L,\mathcal{N}_R) \in \text{SO}(4)_L\times \text{SO}(4)_R \subset \text{SO}(4,4)\,$ by four mass parameters 
\begin{equation}
    \mathcal{N}_L = \begin{pmatrix}
        R(m_1+m_3)&0\\
        0&R(m_1-m_3)
        \end{pmatrix}\,, \qquad \mathcal{N}_R = \begin{pmatrix}
        R(m_2+m_4)&0\\
        0&R(m_2-m_4)
        \end{pmatrix}\,,
        \label{rotations in terms of mass parameters}
\end{equation}
where we use the notation $R(x)=\begin{psmallmatrix}\cos x & \,\,\,\,-\sin x \\ \sin x & \,\,\,\,\cos x \end{psmallmatrix}$ for a $2\times 2$ rotation matrix\footnote{In order to properly define the orbifold action on all the fields, $\mathcal{N}_{L/R}$ should be uplifted to Spin$(4)_{L/R}$; for more details see \cite{Gkountoumis:2023fym}.}. Then, the orbifold acts on the bosonic torus coordinates through (asymmetric) rotations
\begin{equation}\label{orbiaction2}
\begin{aligned}
{W}_{L}^1 \;&\rightarrow\; e^{i(m_1+m_3)}\: {W}_{L}^1 \,, \\
{W}_{L}^2 \;&\rightarrow\; e^{i(m_1-m_3)}\: {W}_{L}^2 \,, \\
W_{R}^1 \;&\rightarrow\; e^{i(m_2+m_4)}\: W_{R}^1 \,, \\
W_{R}^2 \;&\rightarrow\; e^{i(m_2-m_4)}\: W_{R}^2 \,,
\end{aligned}
\end{equation}
and with the same action on their world-sheet superpartners. Symmetric orbifolds arise in the case in which $m_1=m_2$ and $m_3=m_4$. The rotations on the torus are accompanied by a shift along the circle coordinate
\begin{equation}\label{shift}
Z \;\rightarrow\; Z + 2\pi \mathcal{R} / p \,,
\end{equation}
which makes the orbifold freely acting. 

For each mass parameter that is not zero (mod $2\pi$) a pair of gravitini becomes massive. So, for example, in order to obtain an $\mathcal{N}=6$ theory three mass parameters have to be set to zero, while for an $\mathcal{N}=2$ theory only one mass parameter should be set to zero. These mass parameters can be translated to the more familiar language of twist vectors used in the orbifold literature as follows\footnote{Comparing with the notation used in \cite{Gkountoumis:2023fym}, we simply omit here the two first trivial entries of the twist vectors. Comparing with \cite{Baykara:2023plc}, we have $\tilde{u}=\phi_L$ and $u=\phi_R$.}
\begin{equation}
\begin{aligned}
   & \tilde{u} \equiv (\tilde{u}_3,\tilde{u}_4)= \frac{1}{2\pi}\left(m_1+m_3,m_1-m_3\right)\ ,\\
   &u\equiv(u_3,u_4)=\frac{1}{2\pi}\left(m_2+m_4,m_2-m_4\right)\ .
    \end{aligned}
    \label{relation twist vectors-mass parameters}
\end{equation}
With these at hand, it is straightforward to see that
\begin{equation}
    \mathcal{N}_L = \begin{pmatrix}
        R(2\pi\tilde{u}_3)&0\\
        0&R(2\pi\tilde{u}_4)
        \end{pmatrix}\,, \qquad \mathcal{N}_R = \begin{pmatrix}
        R(2\pi u_3)&0\\
        0&R(2\pi u_4)
        \end{pmatrix}\,,
        \label{orbifold twist in terms of twist vectors}
\end{equation}
Finally, we have to ensure that our models satisfy modular invariance. This can be verified if the following conditions hold \cite{vafa1986modular,Baykara:2023plc}
\begin{equation}
    p \sum_{i=3}^4 \tilde{u}_i \,\in\, 2\mathbb{Z} \qquad\text{and}\qquad p \sum_{i=3}^4 u_i \,\in\, 2\mathbb{Z} \ .
    \label{modular conditions on twist vectors}
\end{equation}
Also, if the rank of the orbifold is even, we check the additional condition\footnote{In some cases it is possible to construct consistent orbifolds even if this condition is not satisfied \cite{Harvey:2017rko}.} \begin{equation}\label{modcond2}
  p\mathcal{M}_{\theta}^{p/2}p \equiv  p_L \mathcal{N}_L^{\,p/2} p_L - p_R \mathcal{N}_R^{\,p/2} p_R \,\in\, 2\mathbb{Z}\,. 
\end{equation}
We mention here that the spectra of all the orbifolds discussed in this work can be found by using \autoref{huiberttable} (see appendix \ref{appendix orbifold spectrum}). This table is taken from \cite{Gkountoumis:2023fym} and provides the untwisted orbifold spectrum. This is sufficient for the purposes of this work, in which we focus on the determination of the moduli spaces, since for freely acting orbifolds all fields coming from the twisted sectors are massive for generic values of the circle radius.

\section{Orbifolds, supergravity and symmetry}\label{OSS}

In this section we discuss the 
symmetry structure of the asymmetric orbifold constructions we are considering and in particular examine the
low energy supergravity theory that emerges.

Consider first the theory before  orbifolding, which is the type IIB theory compactified on $T^5$.
The low energy  theory is $D=5$, $\mathcal{N}=8$ supergravity with rigid duality symmetry $G=\text{E}_{6(6)}$ and local R-symmetry $\hat{H}=\text{Sp}(4)$ (sometimes referred to as $\text{USp}(8)$).
The scalar fields take values in the coset $G/H$ where $H$ is the maximal compact subgroup of $\text{E}_{6(6)}$, which is $H=\text{Sp}(4)/\mathbb{Z}_2$. The group $\text{E}_{6(6)}$ has a maximal subgroup Spin$(5,5)\times \text{SO}(1,1)$ which plays a role in what follows. In the string theory, the global $G=\text{E}_{6(6)}$ is broken to its discrete subgroup $G(\mathbb{Z})=\text{E}_{6(6)}(\mathbb{Z})$, forming the U-duality group \cite{Hull:1994ys}. The T-duality subgroup of this is  Spin$(5,5;\mathbb{Z})$.

The orbifolds we are considering are each specified by  a twist matrix which is an element  
$\hat{\cal M}\in  \hat{H}= \text{Sp}(4)$
that projects to an element
$ {\cal M}\in \text{Sp}(4)/\mathbb{Z}_2$ which is in the discrete U-duality subgroup of
$  \text{E}_{6(6)}(\mathbb{Z})$, i.e.\ 
$\mathcal{M}\in  \text{E}_{6(6)}(\mathbb{Z})\cap \text{Sp}(4)/\mathbb{Z}_2$.
We require $\hat{\cal M}$ to satisfy 
$\hat{\cal M}^p=1$ for some $p$, so that it generates a $\mathbb{Z}_p$ group; for more details see \cite{Gkountoumis:2023fym}. We further require that the 
twist is in the 6-dimensional T-duality group arising from compactifying to 6 dimensions on $T^4$. Then
${\cal M}\in$ Spin$(4,4;\mathbb{Z})$ and
$\hat{\cal M}\in$ Spin(4)$\times $Spin(4).
The orbifold we are considering is then by a T-duality twist on $T^4$ combined with a shift on a further circle by $2\pi\mathcal{R}/p$, so that it is a $\mathbb{Z}_p$ orbifold. In more detail, the twist is a T-duality transformation on $T^4$ combined with an R-symmetry twist:
\begin{equation}
    ({\cal M},\hat {\cal M})\in
{\rm Spin}(4,4;\mathbb{Z})
\,
\times \,
[{\rm Spin}(4)\times {\rm Spin}(4) ]
\subset 
\text{E}_{(6)6}(\mathbb{Z})\times \text{Sp}(4)\ .
\end{equation}
In the supergravity theory, the orbifold corresponds to a
Scherk-Schwarz reduction and in string theory is a compactification with duality twist
\cite{Dabholkar:2002sy,Gkountoumis:2023fym}.
In the supergravity theory
the twist 
generates a $\text{U}(1)$ subgroup of 
$\text{E}_{6(6)}\times \text{Sp}(4)$.
Let the subgroup of $\text{E}_{6(6)}\times \text{Sp}(4)$ that commutes with this $\text{U}(1)$ be
\begin{equation}
\mathcal{K}\times \hat{\mathcal{K}} \subset \text{E}_{6(6)}\times \text{Sp}(4)\ .
\end{equation}
Then the twist will break the supergravity
duality symmetry 
and  R-symmetry
but a subgroup
$\mathcal{K}\subset \text{E}_{6(6)}$
of the duality symmetry and a subgroup
$\hat{\mathcal{K}}\subset
\text{Sp}(4)
$
will remain as symmetries.
There will be massless scalar fields
in the coset space
$\mathcal{K}/\bar {\mathcal{K}}$ where 
$\bar{\mathcal{K}}$ is the  compact subgroup of $\mathcal{K}$:
$\bar {\mathcal{K}}=\mathcal{K}\cap \text{Sp}(4)/\mathbb{Z}_2
$.
In string theory, a discrete subgroup
$\mathcal{K}(\mathbb{Z})\subset \text{E}_{6(6)}(\mathbb{Z})$ of the U-duality is expected to remain as a duality symmetry and a local $\hat{\mathcal{K}}\subset \text{Sp}(4)$ as an R-symmetry.

In this paper we  consider orbifolds for which
 the duality group of the effective supergravity theory is a subgroup of Spin$(5,5)\times \text{SO}(1,1)$, which is the aforementioned maximal subgroup of $\text{E}_{6(6)}$.\footnote{Other orbifolds can give rise to  duality symmetries that are subgroups of other maximal subgroups of $\text{E}_{6(6)}$.}
 Here Spin$(5,5)$ arises from the 5-dimensional T-duality group and \text{SO}$(1,1)$ from the 10-dimensional SL(2,$\mathbb{R}$) symmetry. Of course, as we have already discussed, the twist, being in the T-duality group of $T^5$, will further break this group. Then the U-duality group of the effective theory will be given by $\mathcal{K}$, and for our models will be of the form
\begin{equation}
\mathcal{K}=\text{SO}(1,1)\times \mathcal{C} \subset
\text{SO}(1,1)\times
{\rm Spin}(5,5)\,.
\label{ K = so(1,1) x c}
\end{equation}
Here $\mathcal{C}$ is the subgroup of
${\rm Spin}(5,5)$ that commutes with the twist. In the quantum theory we expect the group $\mathcal{C}$ to be broken to its discrete subgroup $\mathcal{C}(\mathbb{Z})$, forming the T-duality group of the orbifold. Moreover, we focus on orbifolds that break the supersymmetry to $\mathcal{N}=2$, so that the orbifold R-symmetry $\hat{\mathcal{K}}$ contains an $\text{SU}(2)$ which is the R-symmetry of the effective $\mathcal{N}=2$ theory that
constitutes the massless sector.

Now, the compact subgroup of $\mathcal{K}$ is
\begin{equation}
\bar {\mathcal{K}}=\mathcal{K}\cap \text{Sp}(4)/\mathbb{Z}_2=\mathbb{Z}_2\times
\bar {\mathcal{C}}
\end{equation}
where
$\bar {\mathcal{C}}=\mathcal{C}\cap 
[{\rm Spin}(5)\times {\rm Spin}(5)]
$.
Then the moduli space is
\begin{equation}
\mathcal{M}_{\mathcal{O}}=\mathcal{K}/\bar {\mathcal{K}} = \mathbb{R}^+
\times \mathcal{C}/\bar {\mathcal{C}}
\subset  \mathbb{R}^+
\times
\frac{
{\rm Spin}(5,5)}
{
{\rm Spin}(5)\times {\rm Spin}(5)
}\,
\end{equation}
 where $\mathbb{R}^+=\text{SO}(1,1)/
\mathbb{Z}_2$.

In this work we will construct models based on asymmetric orbifolds for which the massless spectrum consists purely of NS-NS fields coming only from the untwisted sector. In particular, all fields in the R-R sector, as well as all states in the twisted sectors will be massive. In this case the duality group of the supergravity theory can be at most Spin$(5,5)\times \text{SO}(1,1)$.
The duality group Spin$(5,5)$
only acts on the NS-NS fields
through SO$(5,5)$.
Since we have only massless NS-NS scalars, the   moduli space of the classical effective theory can be at most 
\begin{equation}
\mathcal{M}_{\text{NS}}=\mathbb{R}^+\times \frac{\text{SO}(5,5)}{\text{SO}(5)\times \text{SO}(5)}\,,
\end{equation}
where the $\mathbb{R}^+$ factor is parametrized in terms of the five-dimensional dilaton $\phi_5$ by $\lambda_5=\langle e^{\phi_5}\rangle$,
and the remaining coset is parametrized by the 25 scalars that are the NS-NS moduli  of $T^5$. 
For the orbifold  some of these 25 scalars will become massive and will no longer be moduli of the orbifolded theory. Hence, the 
coset space will
be reduced to a subspace
 parametrized only by those scalars that are invariant under the twist, and as a consequence remain massless and are moduli of the orbifolded theory. On the other hand, the dilaton is inert under the orbifold action and will be a modulus in all of our constructions. The five-dimensional dilaton is T-duality invariant and parametrizes the space $\mathbb{R}^+$ preserved by  $\mathcal{C}$.
 Having said all these, we conclude that the   classical moduli space of the  orbifold   
 $\mathcal{M}_{\mathcal{O}}$ will be exactly the subspace of $\mathcal{M}_{\text{NS}}$ that is compatible with the twist, that is 
 \begin{equation}
\mathcal{M}_{\mathcal{O}}=\mathcal{K}/\bar {\mathcal{K}} = \mathbb{R}^+\times \mathcal{C}/\bar {\mathcal{C}}\,.
\end{equation}
For the quantum theory we expect the U-duality group $\text{SO}(1,1)\times
\mathcal{C}$ to be broken to is discrete subgroup $\mathbb{Z}_2\times\mathcal{C}(\mathbb{Z})$. As we mentioned before, the classical SO(1,1) symmetry is inherited from the type IIB SL(2,$\mathbb{R}$) symmetry in ten dimensions. There, it acts as $\tau \to a^2 \tau$, so that it shifts the (ten-dimensional) dilaton $\Phi$ and rescales the axion $C_0$.\footnote{Recall that $\tau\to\frac{a\tau+b}{c\tau+d}$, where $\tau=C_0+ie^{-\Phi}$.} In the quantum theory, SO(1,1) is broken to $\mathbb{Z}_2$, corresponding to the integer values $a=\pm 1$. This $\mathbb{Z}_2$ leaves the dilaton invariant but acts on the NS-NS and R-R two-forms as $B_2 \to -B_2$ and $C_2\to -C_2$. Upon reducing to five dimension, this $\mathbb{Z}_2$ is preserved since it commutes with the twist, but it does not act on the five-dimensional dilaton. The quantum symmetries and duality groups will play an important role in showing the absence of quantum corrections to the moduli spaces, one of the main results of our paper.

For any string compactification to 5 dimensions that preserves $\mathcal{N}=2$ supersymmetry there is a low-energy effective field theory consisting of $\mathcal{N}=2$ supergravity coupled to hypermultiplets and vector multiplets.
The moduli are the massless scalar fields in the hypermultiplets and vector multiplets, and are governed by non-linear sigma models 
on the hypermultiplet moduli space and vector multiplet moduli space, respectively.

For models such as the ones we are considering here in which the scalar giving the string coupling constant is in a vector multiplet, the hypermultiplet moduli space metric must be independent of the string coupling and so must be given by the classical result with no quantum corrections.
However, in general the vector multiplet moduli space metric may have quantum corrections. 

For our models the supersymmetry is spontaneously broken rather than explicitly broken, and, at least at the classical level, there is a  truncation to an $\mathcal{N}=8, D=5$ supergravity theory.
As we shall see, this means that there is a hidden $\mathcal{N}=8$ local supersymmetry that  relates the hypermultiplets and vector multiplets. We will discuss this in more detail in the next section.

We conclude this section by mentioning that the issue of finding the duality group of the Narain lattice of toroidal orbifolds was first addressed in \cite{Spalinski:1991yw,Spalinski:1991vd}. For some other references see e.g.\ \cite{ferrara1992moduli,Bailin:1993ri,Erler:1992av,{cardoso1994moduli}}.

\section{The effective supergravity theory}\label{EFFSUGRA}

As explained in \cite{Gkountoumis:2023fym}, the  orbifold construction can be viewed as a compactification with duality twist of the kind introduced in \cite{Dabholkar:2002sy}. Type IIB string theory compactified on $T^4$ gives a 6-dimensional theory with U-duality group $\text{Spin}(5,5,\mathbb{Z})$ and
scalars taking values in the moduli space of IIB on $T^4$, which is $\text{Spin}(5,5)/\left[\text{Spin}(5)\times \text{Spin}(5)\right]$
quotiented by the U-duality group \cite{Hull:1994ys}.
This $6D$ theory is then
compactified on a further circle with a U-duality twist, i.e.\ with a  $\text{Spin}(5,5,\mathbb{Z})$ monodromy M on the circle.
At a point in the moduli space that is fixed under the action of M, the construction becomes an orbifold by the symmetry M, which is a symmetry of the IIB theory on  $T^4$ at that point in moduli space, combined with a shift on the extra circle \cite{Dabholkar:2002sy}.  For the models considered here the monodromy is a T-duality transformation, i.e.\ $\text{M}=\mathcal{M}$, so that this construction gives a freely acting  orbifold. This viewpoint is useful as it enables us to formulate the construction without restricting to a particular point in moduli space.
 
 The toroidal compactification of IIB supergravity to 6 dimensions has a consistent truncation to 6-dimensional $\mathcal{N}=8$ supergravity, and this is also a consistent truncation of the full string theory compactified on $T^4$. Then the string theory compactification with duality twist has a consistent truncation to  the Scherk-Schwarz compactification of
$\mathcal{N}=8$ supergravity in 6 dimensions on $S^1$.
This compactification 
gives a gauged $\mathcal{N}=8$ supergravity in 5 dimensions
coupled to
an infinite number of massive
$\mathcal{N}=8$ supermultiplets arising from the twisted circle reduction.
This $\mathcal{N}=8$ theory
in turn has a consistent truncation to
 give an $\mathcal{N}=2$ supergravity coupled to supermatter in 5 dimensions, resulting from  the massless fields that are invariant under the twist \cite{Duff:1985jd}. 
This is in accord with the arguments of \cite{Lee:2014mla}.

Of particular interest are the cases in which there is in fact a consistent truncation to a
 gauged $\mathcal{N}=8$ supergravity. We believe this to be the case for all of our examples here, but we will not assume this. Instead, we shall show that our results are consistent with this.

 In some cases there may be additional \say{accidental} massless scalars of the type discussed in \cite{Hull:2020byc} (cf. section 3.9).  Such accidental modes arise when some (or all) of the mass parameters add up to $2\pi n$, $n \in \mathbb{Z}^*$.
 Then some of the massive $\mathcal{N}=8$ supermultiplets arising from the twisted circle reduction 
 contain massless scalars. In particular, these $\mathcal{N}=8$ supermultiplets
 decompose into $\mathcal{N}=2$ supermultiplets of different masses and in the accidental case some of these $\mathcal{N}=2$ supermultiplets turn out to be massless.
 For the rest of this section, we restrict ourselves to the cases in which there are no accidental massless scalars of this type.

The scalars of the  $\mathcal{N}=8$ supergravity multiplet
include the moduli of the $T^4$
and
have a scalar potential.
The scalar potential   has a minimum at which the potential vanishes and the location of the minimum in the scalar target space fixes the $T^4$ moduli to be at the point where the $T^4$ CFT is invariant under the $\mathbb{Z}_p$ symmetry that we use in the orbifold. At the  minimum there is a Minkowski vacuum  in which
the $\mathcal{N}=8$ supersymmetry is spontaneously broken. For the models we consider here, it is broken  to $\mathcal{N}=2$ and then the $\mathcal{N}=8$ supergravity multiplet can be decomposed into  $\mathcal{N}=2$ multiplets.
The massless $\mathcal{N}=2$  multiplets give $\mathcal{N}=2$ supergravity coupled to hypermultiplets and vector multiplets. In addition, there are massive multiplets including $\mathcal{N}=2$ gravitino multiplets that contain 6 massive gravitini. If the mass parameters $m_i$ in the twist are set to zero, the massive multiplets all become massless and ungauged $\mathcal{N}=8, D=5$ supergravity is recovered.

The scalars of the ungauged
$\mathcal{N}=8$ supergravity in 5 dimensions take values in the coset space $\text{E}_{6(6)}/\left[\text{Sp}(4)/\mathbb{Z}_2\right]
$. Our construction results in some of these scalars being eaten by vector fields which become massive and some gaining masses through the scalar potential.
The remaining scalars which are massless are the moduli. They correspond to flat directions of the potential and   lie in a subspace of
$\text{E}_{6(6)}/\left[\text{Sp}(4)/\mathbb{Z}_2\right]
$.
As we argued in the previous section, these remaining scalars 
in fact take values in the classical moduli space
\begin{equation}
\mathcal{M}_{\mathcal{O}}=\mathcal{K}/\bar {\mathcal{K}} = \mathbb{R}^+\times \mathcal{C}/\bar {\mathcal{C}}\,.
\end{equation}
For compactifications with 
$\mathcal{N}=2$ supersymmetry this space must factorise into a vector multiplet moduli space and 
a hypermultiplet moduli space
\begin{equation}
\mathcal{M}_{\mathcal{O}}=
{\cal M}_V \times {\cal M}_H\,,
\end{equation}
so that $\mathcal{C}$ 
must have a factorisation
\begin{equation}
\mathcal{C}=\mathcal{C}_V
\times
\mathcal{C}_H\,,
\end{equation}
with a corresponding 
factorisation of
$\bar {\mathcal{C}}$
\begin{equation}
\bar {\mathcal{C}}=\bar {\mathcal{C}}_V
\times
\bar {\mathcal{C}}_H\,,
\end{equation}
so that
\begin{equation}
{\cal M}_V 
=\mathbb{R}^+\times 
{\mathcal{C}}_V/\bar {\mathcal{C}}_V\qquad\text{and} \qquad {\cal M}_H 
=
{\mathcal{C}}_H/\bar {\mathcal{C}}_H\,.
\label{factorization of C}
\end{equation}
The metric on the moduli space 
$\mathcal{M}_{\mathcal{O}}=
{\cal M}_V \times {\cal M}_H$
is the restriction of the  
canonical coset metric
on
the $\text{E}_{6(6)}/\left[\text{Sp}(4)/\mathbb{Z}_2\right]
$ coset space to this subspace.

In the quantum theory,  the set of moduli remains the same (since we are restricting to the cases with no extra accidental massless scalars) but the  metric 
for the vector moduli space ${\cal M}_V$
could in principle receive quantum corrections.
However, 
$\mathcal{N}=8$ local supersymmetry should be preserved in the quantum theory and this is highly restrictive.
The quantum metric must also be a
restriction of the
coset metric
on
the $\text{E}_{6(6)}/\left[\text{Sp}(4)/\mathbb{Z}_2\right]$.
One way in which this could happen that is consistent with the symmetries of the system  would be if 
${\mathcal{C}}$ were deformed to a conjugate subgroup
\begin{equation}
{\mathcal{C}}'=g\,{\mathcal{C}}g^{-1}
\end{equation}
where
$g\equiv g(\phi_5) $
is a dilaton-dependent element of $\text{E}_{6(6)}$. This transformation would need to preserve
${\cal M}_H $ as this receives no quantum corrections, so 
$g\,{\mathcal{C}}_Hg^{-1}={\mathcal{C}}_H$.
However, such a deformation would also lead to a deformation of the twist ${\cal M}$
 and, since this deformation depends continuously on the dilaton, will be inconsistent with the 
 restriction that
${\cal M}\in$ Spin$(4,4;\mathbb{Z})$.

It seems that any deformation would be ruled out in this way.
This would then imply that the metric on the vector multiplet moduli space can receive no quantum corrections. In section 7, we shall give independent arguments that this is the case, using more details of the models under consideration.
These arguments also extend to the case with accidental massless scalars.
This then provides further evidence supporting the discussion in this section.

\section{$\mathcal{N}=2$ models in $5D$ with exact moduli spaces}
\label{N=2 models in 5d with exact moduli spaces}

In this section the idea is to construct models where there are no massless fields coming from the R-R sector, and all fields from the twisted sectors are massive. Such models can be realised via freely acting asymmetric orbifolds. If there are no massless R-R scalars, the moduli space of both vector multiplets and hypermultiplets is entirely spanned by NS-NS scalars. Since there are no massless R-R vectors, there are no D-branes carrying R-R charges that can correct the metric on the moduli space, but there are presumably non-supersymmetric D-branes. Then the classical moduli space can be determined as described in section \ref{OSS}. Note that if the orbifold spectrum contains massless states coming from the twisted sectors, the moduli space will be enhanced. However, our orbifolds are freely acting and all the states coming from the twisted sectors are massive for generic values of the circle radius. In all the examples we discuss here, we will take the radius of the circle to be large enough compared with the string scale to ensure that there will be no massless states coming from the twisted sectors (see also \cite{Baykara:2023plc}). 

Notice that our analysis of the moduli space is exact in $\alpha'$. This is because we have determined the T-duality group $\mathcal{C}$ exactly. Any quantum correction in $\alpha'$ should come with a dimensionful parameter, such as e.g.\ $L/\sqrt{\alpha'}$ for any radius $L$ on the $T^5$. Such corrections would not be compatible with the symmetry $\mathcal{C}$ acting on the moduli space $\mathcal{C}/\mathcal{C}_H$. As in the case of higher extended supersymmetry $\mathcal{N}\geq 4$, $\sqrt{\alpha'}$ corrections will however contribute to higher derivative terms in the effective action.

\subsection{Determination of the moduli space}

In order to determine the moduli space of the orbifold we need to find the group $\mathcal{C}\subset \text{Spin}(5,5)$ that commutes with the twist. For models with only NS-NS scalars, it is sufficient to work with the subgroup $\text{SO}(5,5)\subset \text{Spin}(5,5)$, as explained in section \ref{OSS}. A general element $\hat{h}\in \text{SO}(5,5)$ is a $10\times 10$ matrix
\begin{equation} \hat{h}=
   \begin{pmatrix}
\hat{a}&\hat{b}\\
\hat{c}&\hat{d}
\end{pmatrix}
\label{general SO(5,5) element}
\end{equation}
that satisfies
\begin{equation}
    \hat{h}^t\, \eta\,\hat{h} = \eta\,,  \qquad \eta=\begin{pmatrix}
    {1}_5 & 0\\
0& -1_5 
\end{pmatrix}\,, 
\label{eta frame}
\end{equation}
and $\text{det}(\hat{h})=1$. Here $\hat{a},\hat{b},\hat{c},\hat{d}$ are $5\times 5$ matrices, $t$ denotes the transpose matrix and $1_5$ is the $5\times 5$ unit matrix. From \eqref{general SO(5,5) element},\eqref{eta frame} we see that the matrices $\hat{a},\hat{b},\hat{c},\hat{d}$ should satisfy the constraints
\begin{equation}
    \hat{a}^t\hat{a}- \hat{c}^t\hat{c}=1_5\,,\qquad   \hat{a}^t\hat{b}- \hat{c}^t\hat{d}=0_5\,,\qquad \hat{d}^t\hat{d}- \hat{b}^t\hat{b}=1_5\,.
    \label{matrices a,b,c,d constraints}
\end{equation}
We will also express the orbifold action in terms of an $\text{SO}(5,5)$ matrix. We embed $\mathcal{M}_{\theta}=(\mathcal{N}_L,\mathcal{N}_R) \in \text{SO}(4)_L\times \text{SO}(4)_R \subset \text{SO}(4,4)\,$ in $\text{SO}(5,5)$ as follows
\begin{equation}
    \hat{\mathcal{M}}_{\theta}=(\hat{\mathcal{N}}_L,\hat{\mathcal{N}}_R) \in \text{SO}(5)_L\times \text{SO}(5)_R \subset \text{SO}(5,5)\,,
    \label{general twist element}
\end{equation}
with
\begin{equation}\hat{\mathcal{N}}_L=
     \begin{pmatrix}
\mathcal{N}_L&0\\
0&1
\end{pmatrix}\,, \qquad \hat{\mathcal{N}}_R=
     \begin{pmatrix}
\mathcal{N}_R&0\\
0&1
\end{pmatrix}\,.
\label{general twist embedded in SO(5,5)}
\end{equation}
Now, in order to find the matrix $\hat{h}$ that commutes with the twist, we simply need to solve the matrix equation
\begin{equation}
    \left[\hat{\mathcal{M}}_{\theta},\hat{h}\right]=0\,.
    \label{commutant}
\end{equation}
This gives the following set of equations
\begin{equation}
  \left[\hat{\mathcal{N}}_L,\hat{a}\right]=0\,,\qquad  \hat{\mathcal{N}}_L\,\hat{b}=\hat{b}\,\hat{\mathcal{N}}_R\,,\qquad  \hat{\mathcal{N}}_R\,\hat{c}=\hat{c}\,\hat{\mathcal{N}}_L\,,\qquad \left[\hat{\mathcal{N}}_R,\hat{d}\right]=0\,.  
  \label{general commutation relations for h,g}
\end{equation}
In practise, in order to solve \eqref{commutant}, we start from a given twist $\hat{\mathcal{M}}_{\theta}\in \text{SO}(5,5)$ and a general matrix $\hat{h}=\begin{psmallmatrix}
\hat{a}&\hat{b}\\
\hat{c}&\hat{d}
\end{psmallmatrix}\in \text{GL}(10;\mathbb{R})$. Then we specify the form of the sub-matrices $\hat{a},\hat{b},\hat{c},\hat{d}$ satisfying \eqref{general commutation relations for h,g}. Afterwards, we demand that $\hat{h}\in \text{SO}(5,5)$ by imposing the conditions \eqref{matrices a,b,c,d constraints} and det$(\hat{h})=1$. This yields the group ${\mathcal{C}}$ that commutes with the twist. Then, as we discussed in section \ref{OSS}, the moduli space will be $\mathbb{R}^+\times \mathcal{C}/\bar {\mathcal{C}}$, where $\bar {\mathcal{C}}$ is the compact subgroup of ${\mathcal{C}}$. 

In what follows we will apply this general analysis to specific examples with $\mathcal{N}=2$ supersymmetry in $D=5$. In order to construct such models we need to turn on three mass parameters. To start with, we choose $m_1,m_2,m_3\neq 0$ (mod $2\pi$) and $m_4=0$. Then, as can be read off from \autoref{huiberttable}, models without massless R-R states should satisfy the following conditions
\begin{equation}
    \begin{aligned}
        &\pm m_1 \pm m_2 \neq 0\quad \text{mod} \quad 2\pi\,,\\
         &\pm m_3 \pm m_2 \neq 0\quad \text{mod} \quad 2\pi\,,
    \end{aligned}
    \label{mass parameters for no R-R fields}
\end{equation}
for any choice of signs. We could have also chosen another vanishing mass parameter instead of $m_4$, e.g.\ $m_1=0$. This change would lead to dual models; these will be discussed in section \ref{dual pairs}. In this section, we  work with $m_4=0$. Using the results of \cite{Gkountoumis:2023fym,Baykara:2023plc}, we can provide a few concrete examples of models with no R-R massless states. In the next subsections we will present such examples with $0,1$ or $2$ hypermultiplets, which involve orbifolds of rank $4,6$ and $12$. We mention here that we did not find any models of rank $2$ and $3$ without massless R-R states. 

\subsection{Models with no hypermultiplets}

In this subsection we focus on models with no hypermultiplets, $n_H=0$, so there will only be vector multiplets. These are quite special in the landscape of string theories, as such models cannot be obtained from geometric compactifications such as M-theory on a CY$_3$. Using \autoref{huiberttable}, we see that such models can be obtained if 
\begin{equation}\label{nohyper}
    \pm m_1 \pm m_2 \pm m_3\neq 0\quad \text{mod} \quad 2\pi\,,
\end{equation}
for any choice of signs. We mention here that each choice of signs that does not satisfy the above condition results in two  massless scalars (and one massless dilatino). Such choices are two-fold degenerate, so we obtain one massless hypermultiplet  for each choice of signs for which $\pm m_1 \pm m_2 \pm m_3$ vanishes.

We will present three  models with  $n_H=0$. These are models studied in \cite{Gkountoumis:2023fym} and \cite{Baykara:2023plc}; for more details we refer to these papers. Some of the models with $n_H=0$ presented in \cite{Gkountoumis:2023fym,Baykara:2023plc} contain both R-R and NS-NS fields. Here we will consider only models with massless spectrum consisting purely of NS-NS fields. As there are no hypermultiplets, these massless scalars will define the vector multiplet moduli space. Our goal is then to determine the classical vector multiplet moduli space (which contains the dilaton). In the subsequent sections we will argue that this moduli space is exact.

\subsubsection{$\mathbb{Z}_{12}$ with $n_V=2$ and $n_H=0$}
\label{z12 with nv=2 and nh=0}

The first example we consider is the orbifold with the following mass parameters
\begin{equation}
    m_1=\frac{2\pi}{3}\ , \qquad m_2= \frac{\pi}{2}\ ,\qquad m_3=\frac{\pi}{3} \ ,\qquad m_4=0\ .
\end{equation}
The corresponding twist vectors are
\begin{equation}
    \tilde{u}=\left(\frac{1}{2},\frac{1}{6}\right)\ ,\qquad u=\left(\frac{1}{4},\frac{1}{4}\right)\,,
    \label{twist vectors for primary example}
\end{equation}
which satisfy \eqref{modular conditions on twist vectors} (for this model $p=12$). These vectors give rise to the twist matrices
\begin{equation}
    \mathcal{N}_L=\begin{pmatrix}
        R(\pi)&0\\
        0&R(\frac{\pi}{3})
    \end{pmatrix}\,,\qquad \mathcal{N}_R=\begin{pmatrix}
        R(\frac{\pi}{2})&0\\
        0&R(\frac{\pi}{2})
    \end{pmatrix}\,.
    \label{twist matrices for primary example}
\end{equation}
This is one of the models constructed in \cite{Baykara:2023plc}. The massless spectrum of this orbifold consists of the $\mathcal{N}=2, D=5$ gravity multiplet and two vector multiplets.
The appropriate lattice to begin with is
\begin{equation}
    \Gamma^{5,5}=\Gamma^{4,4}(D_4)\oplus\Gamma^{1,1}\ .
    \label{lattice d4 + gamma1,1}
\end{equation}
Given the twist matrices \eqref{twist matrices for primary example} and using the properties of the $D_4$ lattice\footnote{One such useful property is that for $p_L-p_R \in \Lambda_R(D_4)$, $p_L^2+p_R^2\in 2\mathbb{Z}$.}, it can be verified that \eqref{modcond2} is satisfied. Note that since the orbifold acts non trivially in all torus directions, the invariant sublattice is simply $\Gamma^{1,1}$.  

Let us now proceed with the calculation of the commutant. In order to find the matrix $\hat{h}=\begin{psmallmatrix}\hat{a} & \hat{b} \\ \hat{c} & \hat{d} \end{psmallmatrix}$, we need to solve \eqref{general commutation relations for h,g} for the twist specified by \eqref{twist matrices for primary example}. We find
\begin{equation}
    \hat{a}=\begin{pmatrix}
           a_{4\times 4} &0_{4\times 1}\\
            0_{1\times 4} &  a_{1\times 1}
        \end{pmatrix}\,,\qquad \hat{b}=\begin{pmatrix}
           0_{4\times 4} &0_{4\times 1}\\
            0_{1\times 4} &  b_{1\times 1} 
        \end{pmatrix}\,,\qquad \hat{c}=\begin{pmatrix}
           0_{4\times 4} &0_{4\times 1}\\
            0_{1\times 4} &  c_{1\times 1}
        \end{pmatrix}\,,\qquad  \hat{d}=\begin{pmatrix}
           d_{4\times 4} &0_{4\times 1}\\
            0_{1\times 4} &  d_{1\times 1}
        \end{pmatrix}\,,
\end{equation}
where $a_{1\times 1},b_{1\times 1},c_{1\times 1},d_{1\times 1}$ are arbitrary numbers, $a_{4\times 4}$ takes the form
\begin{equation}a_{4\times 4}=
    \begin{pmatrix}
    a^1_{2\times 2} & 0_{2\times 2}\\
           0_{2\times 2} & a^2_{2\times 2}\\
\end{pmatrix}\,,\quad \text{with} \quad
a^2_{2\times 2}=
    \begin{pmatrix}
      a & \tilde{a}\\
      -\tilde{a} & a
    \end{pmatrix}\,,\quad\text{and}\quad  a^1_{2\times 2}\quad \text{arbitrary}\,,
    \label{matrix a form commuting}
\end{equation}
and $d_{4\times 4}$ is of the following form
\begin{equation}d_{4\times 4}=
\begin{pmatrix}
    d^1_{2\times 2} & d^2_{2\times 2}\\
            d^3_{2\times 2} & d^4_{2\times 2}\\
\end{pmatrix}\,,\quad \text{with} \quad
d^i_{2\times 2}=
    \begin{pmatrix}
      d_i & \tilde{d}_i\\
      -\tilde{d}_i & d_i
    \end{pmatrix}\,,\quad i=1,\ldots4\,.
    \label{d SO(4) subgroup}
\end{equation}
It will be useful to rearrange some columns and rows of $\hat{h}$ in order to bring $d_{4\times 4}$ to the form\footnote{This rearrangement doesn't affect the other sub-matrices in $\hat{h}$. Also, a permutation of rows and columns is an automorphism of $\text{GL}(10;\mathbb{R})$.} (see also \cite{cardoso1994moduli} for more details)
\begin{equation}
      \begin{pmatrix}
      d_1&d_2&\tilde{d}_1&\tilde{d}_2\\
      d_3&d_4&\tilde{d}_3&\tilde{d}_4\\
       -\tilde{d}_1&-\tilde{d}_2&{d}_1&{d}_2\\
       -\tilde{d}_3&-\tilde{d}_4&{d}_3&{d}_4
    \end{pmatrix} \,.
\end{equation}
Now, we can use the isomorphism
\begin{equation}
    \begin{pmatrix}
      x & \tilde{x}\\
      -\tilde{x} & x
    \end{pmatrix}\,\in \text{GL}(2n;\mathbb{R})\, \cong \, z =x+i\tilde{x} \in   \text{GL}(n;\mathbb{C})\,,
    \label{GL isomorphism}
\end{equation}
where $x$ and $\tilde x$ are real $n\times n$ matrices, in order to represent $d_{4\times 4}$ as
\begin{equation}
 \begin{pmatrix}
      d_1&d_2&\tilde{d}_1&\tilde{d}_2\\
      d_3&d_4&\tilde{d}_3&\tilde{d}_4\\
       -\tilde{d}_1&-\tilde{d}_2&{d}_1&{d}_2\\
       -\tilde{d}_3&-\tilde{d}_4&{d}_3&{d}_4
    \end{pmatrix} \xrightarrow{\eqref{GL isomorphism}}  \begin{pmatrix}
      d_1+i\tilde{d}_1 &  d_2+i\tilde{d}_2\\
       d_3+i\tilde{d}_3 &  d_4+i\tilde{d}_4
    \end{pmatrix}= \begin{pmatrix}
      z_1 & z_2\\
      z_3 & z_4
    \end{pmatrix} \in   \text{GL}(2;\mathbb{C})\,.
    \label{d4 matrix complex rep}
\end{equation}
So far, we have found the form of a generic matrix $\hat{h}\in \text{GL}(10;\mathbb{R})$ that commutes with the particular twist specified by \eqref{twist matrices for primary example}. Now, we demand that $\hat{h}\in \text{SO}(5,5)$. First, by imposing the constraints \eqref{matrices a,b,c,d constraints} we find
    \begin{align}
       & {a}_{4\times 4}^t{a}_{4\times 4}=1_{4\times 4}\,,\label{aal1}\\
       &{a}_{1\times 1}^t{a}_{1\times 1}-c_{1\times 1}^tc_{1\times1}=1_{1\times 1}\,,\label{aal2}\\
       & {a}_{1\times 1}^tb_{1\times 1}-c_{1\times 1}^td_{1\times 1}=0_{1\times 1}\,,\label{aal3}\\
       &d_{1\times 1}^td_{1\times 1}-b_{1\times 1}^tb_{1\times 1}=1_{1\times1}\,,\label{aal4}\\
       &d_{4\times 4}^td_{4\times 4}=1_{4\times4}\,.\label{aal5}
          \end{align}
From \eqref{aal1} and \eqref{matrix a form commuting} we see that $a^i_{2\times 2}\in \text{O}(2)$, for $i=1,2$. From \eqref{aal2}-\eqref{aal4} it follows that $a_{1\times 1},b_{1\times 1},c_{1\times 1}$ and $d_{1\times 1}$ form an $\text{O}(1,1)$ group. Also, \eqref{aal5} together with \eqref{d4 matrix complex rep} yield $d_{4\times 4}\in \text{U}(2)\subset \text{O}(4)$. Finally, by demanding det$(\hat{h})=1$ we conclude that the group commuting with the twist is
\begin{equation}
   \mathcal{C}= \text{SO}(1,1)\times \text{SO}(2)^2 \times \text{SU}(2)\times \text{U}(1)\,.
\end{equation}
As explained in section \ref{OSS}, the moduli space is given by
\begin{equation}
    \mathbb{R}^+\times \mathcal{C}/ \bar {\mathcal{C}}\,.
\end{equation}
Hence, we find 
\begin{equation}\label{modspace2}
    {\cal M}_V=\mathbb{R}^+\times \mathbb{R}^+\ ,
\end{equation}
where we recall that $\mathbb{R}^+=\text{SO}(1,1)/\mathbb{Z}_2$. This moduli space is parametrized by the string coupling and the radius of the circle $\mathcal{R}$. Also, the  isometry group of the moduli space is $\text{SO}(1,1)\times \text{SO}(1,1)$. 

In terms of real special geometry \cite{Gunaydin:1983bi,Gunaydin:1983rk}, the moduli space \eqref{modspace2} can be described by the cubic polynomial
\begin{equation}
    C(h)=d_{ABC}h^Ah^Bh^C\ ,
\end{equation}
with the only two non-vanishing $d$-symbols being\footnote{The overall normalization of the $d$-symbols is not fixed by the arguments here, but fixed by the conventions used in e.g.\ \cite{deWit:1991nm}, see eqn. (1.13) of that reference. If we keep the normalization arbitrary, but with $d_{133}=-d_{122}$, then the metric in \eqref{classmetric} is multiplied by $d_{122}$ which has to be taken positive for positive definiteness of the metric.}
\begin{equation}\label{d-symbols1}
    d_{122}=1\ ,\qquad d_{133}=-1\ .
\end{equation}
This gives
\begin{equation}
    C(h)=3h^1\Big((h^2)^2-(h^3)^2\Big)\,.
    \label{cubic pol for ex1}
\end{equation}
This is a special example of a homogeneous real special geometry. It falls into the classification of $d$-symbols given in \cite{deWit:1991nm, deWit:1992wf} corresponding to homogeneous real special geometries. The moduli spaces of real special geometry are found by setting $C(h)=1$. For the cubic polynomial \eqref{cubic pol for ex1}, this equation can be solved  by the parametrization
\begin{equation}\label{repsh}
    h^1=\frac{1}{3a^2}e^{2\varphi/3}\ ,\qquad h^2= a\,e^{-\varphi/3}\cosh{(\sigma/\sqrt{3})}\ ,\qquad h^3=a\,e^{-\varphi/3}\sinh{(\sigma/\sqrt{3})}   \ ,
\end{equation}
for any real constant $a$. The induced metric can then be computed from
\begin{equation}\label{classmetric}
    {\rm d}s^2=-9\,d_{ABC}h^A{\rm d}h^B{\rm d}h^C= {\rm d}\varphi^2+{\rm d}\sigma^2   \ ,
\end{equation}
which is the canonical, flat metric on \eqref{modspace2}. The normalization is chosen such that later we can identify the expectation value of $e^{\varphi}$ with the (five-dimensional) string coupling and the expectation value of $e^{2\sigma/\sqrt{3}}$
with 
$\mathcal{R}/\sqrt{\alpha'}$ where $\mathcal{R}$ is the radius of the circle.

Now, the duality group of the effective theory, $\mathcal{K}=\text{SO}(1,1)\times \mathcal{C}$, contains $\text{SO}(1,1)\times \text{SO}(1,1)$. One SO(1,1) isometry group acts linearly on $(h^2,h^3)$ and leaves the cubic polynomial invariant. The part connected to the identity corresponds to a shift in $\sigma$, so is  a rescaling of the radius $\mathcal{R}$. The SO(1,1) element $-1_{2\times 2}$ is realised by $a\to -a$, i.e.\ $(h^2,h^3)\to -(h^2,h^3)$, leaving the radius invariant. The other continuous SO(1,1) isometry is a scale transformation generating  a shift in the dilaton, see also section \ref{OSS}. This acts as 
$h^1\to e^{-2b}h^1, h^2\to e^b h^2, h^3\to e^b h^3$. This does not act on the radius $\mathcal{R}$ and therefore the two SO(1,1) factors commute. In the quantum theory we expect that both SO(1,1) factors are broken to $\mathbb{Z}_2$ subgroups under which the dilaton is invariant (see section \ref{OSS}), while $(h^2,h^3)\to -(h^2,h^3)$. We discuss this further in section \ref{absence}.

\subsubsection{$\mathbb{Z}_{12}$ with $n_V=4$ and $n_H=0$}
\label{Z12 example with zero hypers}

As a second example, let us consider the orbifold with mass parameters 
\begin{equation}
    m_1=\frac{2\pi}{3}\ , \qquad m_2= \frac{\pi}{2}\ ,\qquad m_3=\frac{2\pi}{3} \ ,\qquad m_4=0\ ,
\end{equation}
or, in terms of the twist vectors, 
\begin{equation}
    \tilde{u}=\left(\frac{2}{3},0\right)\ ,\qquad u=\left(\frac{1}{4},\frac{1}{4}\right)\ .
    \label{twist vectors z12 4 vm}
\end{equation}
This model was also discussed in \cite{Baykara:2023plc}, and the appropriate lattice  is \begin{equation}
    \Gamma^{5,5}=\Gamma^{4,4}(D_4)\oplus\Gamma^{1,1}\ .
\end{equation}
The conditions for modular invariance can be checked similarly with the previous example. The massless spectrum of this orbifold consists of the $\mathcal{N}=2, D=5$ gravity multiplet and four vector multiplets.

To find the invariant sublattice $I\subset \Gamma^{4,4}(D_4)$  we work in two steps. First, we see from the twist vector $u$ in \eqref{twist vectors z12 4 vm} that $p_R={0}$, which, combined with the condition $p_L-p_R \in \Lambda_R({D_4})$, implies that $p_L\in \Lambda_R({D_4})$, that is $p_L=(p_L^1,p_L^2,p_L^3,p_L^4)$, with $p_L^i\in \mathbb{Z}$ and $\sum_i p_L^i \in 2\mathbb{Z}$, for $i=1,\ldots,4$.\footnote{In general $\Lambda_R({D_n})$ is the lattice of vectors with integer entries, summing up to an even number.} Secondly, from the twist vector $\tilde{u}$ in \eqref{twist vectors z12 4 vm} we see that not all components of $p_L$ are invariant under the orbifold action, which indicates that the invariant lattice is spanned by the vectors $(p_L,0)=(0,0,p_L^3,p_L^4,{0})$, with $p_L^3,p_L^4\in \mathbb{Z}$ and $p_L^3+p_L^4\in2\mathbb{Z}$, namely $I=\Lambda_R({D_2})$. Finally, recall that the orbifold leaves $\Gamma^{1,1}$ invariant, and hence the complete orbifold invariant lattice is
\begin{equation}
    \hat{I}=I \oplus\Gamma^{1,1}\ .
\end{equation}
We now move on to the calculation of $\mathcal{C}$. For the twist specified by \eqref{twist vectors z12 4 vm} the equations \eqref{general commutation relations for h,g} can be solved by the following matrices
\begin{equation}
     \hat{a}=\begin{pmatrix}
           a_{2\times 2} &0_{2\times 3}\\
            0_{3\times 2} &  a_{3\times 3}
            \end{pmatrix}\,,\qquad\hat{b}=\begin{pmatrix}
           0_{4\times 4} &0_{2\times 1}\\
            0_{1\times 4} &  b_{3\times 1} \end{pmatrix}\,,\qquad\hat{c}=\begin{pmatrix}
           0_{4\times 4} &0_{4\times 1}\\
            0_{1\times 2} &  c_{1\times 3} \end{pmatrix}\,,\qquad  \hat{d}=\begin{pmatrix}
           d_{4\times 4} &0_{4\times 1}\\
            0_{1\times 4} &  d_{1\times 1}
        \end{pmatrix}\,,
        \label{z12 matrices a,b,c,d}
\end{equation}
where $a_{3\times 3},b_{3\times 1},c_{1\times 3}, d_{1\times 1}$ are arbitrary, $a_{2\times 2}$ takes the same form as the matrix $a_{2\times 2}^2$ in \eqref{matrix a form commuting} and $d_{4\times 4}$ is the same as in \eqref{d SO(4) subgroup}. As in the previous example, we use the isomorphism \eqref{GL isomorphism} in order to rewrite $d_{4\times 4}$ as in \eqref{d4 matrix complex rep}. Then, imposing the constraints \eqref{matrices a,b,c,d constraints} yields
 \begin{align}
       & {a}_{2\times 2}^t{a}_{2\times 2}=1_{2\times 2}\,,\label{aaal1}\\
       &{a}_{3\times 3}^t{a}_{3\times 3}-c_{3\times 1}^tc_{1\times3}=1_{3\times 3}\,,\label{aaal2}\\
       & {a}_{3\times 3}^tb_{3\times 1}-c_{3\times 1}^td_{1\times 1}=0_{3\times 1}\,,\label{aaal3}\\
       &d_{1\times 1}^td_{1\times 1}-b_{1\times 3}^tb_{3\times 1}=1_{1\times1}\,,\label{aaal4}\\
       &d_{4\times 4}^td_{4\times 4}=1_{4\times4}\,.\label{aaal5}
          \end{align}
From \eqref{aaal1} and  \eqref{matrix a form commuting} we see that $a_{2\times 2}\in \text{O}(2)$. From \eqref{aaal2}-\eqref{aaal4} it follows that $a_{3\times 3},b_{3\times 1},c_{1\times 3}$ and $d_{1\times 1}$ form an $\text{O}(3,1)$ group. Also, \eqref{aaal5} together with \eqref{d4 matrix complex rep} yield $d_{4\times 4}\in \text{U}(2)\subset \text{O}(4)$. Finally, imposing $\text{det}(\hat{h})=1$ yields
\begin{equation}
   \mathcal{C}=\text{SO}(3,1)\times \text{SO}(2) \times \text{SU}(2)\times \text{U}(1)\,.
\end{equation}
Now, the moduli space can be easily determined by the analysis of section \ref{OSS}. We find
\begin{equation}
     {\cal M}_V=\mathbb{R}^+\times \frac{\text{SO}(3,1)}{\text{SO}(3)}\ ,
\end{equation}
and we recall that $\mathbb{R}^+$ is parametrized by the string coupling.

So far, we have constructed a seemingly consistent string theory, with a modular invariant partition function and a moduli space consistent with $\mathcal{N}=2, D=5$ local supersymmetry.
However, this model seems to fail at satisfying the integrality condition that is related to the degeneracies of states in the twisted sectors; models with the same issue were discussed recently in \cite{bianchi2022perturbative,Gkountoumis:2023fym}. In particular, the degeneracy of the ground state in the twisted sectors $D(k)$ is given by the number of \say{chiral} fixed points $\tilde{\chi}\cdot\chi$ divided by the volume of the invariant sublattice $\text{Vol}(\hat{I})$, and for a consistent physical theory $D(k)$ should be an integer \cite{Narain:1986qm,narain1991asymmetric}. Now, we can compute for example $D(k=1)$ for the $\mathbb{Z}_{12}$ orbifold discussed above. We find $\tilde{\chi}\cdot\chi=2 \sin \left(\frac{2\pi}{3}\right) 4\sin^2\left(\frac{\pi}{4}\right)= 2\sqrt{3}$ and $\text{Vol}(\hat{I})=2$, which implies $D(k=1)=\sqrt{3}$.

On the basis of this, such a model would usually be discarded. However, in section \ref{dual pairs} we will construct a dual model (see the first model in \autoref{Table of dual pairs with R-R field}), which does not suffer from the same problem. So, it would be interesting to see if any physical interpretation  could be given to the non-integral degeneracy, or if the dual model should also be discarded, presumably due to a non-perturbative inconsistency.

\subsubsection{$\mathbb{Z}_6$ with $n_V=6$ and $n_H=0$}
\label{{Z}_6 with n_V=6 and n_H=0}

As a final example with no hypermultiplets, consider the orbifold with mass parameters
\begin{equation}
    m_1=\pi\ , \qquad m_2= \frac{2\pi}{3}\ ,\qquad m_3=\pi \ ,\qquad m_4=0\ .
\end{equation}
The corresponding twist vectors are
\begin{equation}
    \tilde{u}=\left(1,0\right)\,,\qquad u=\left(\frac{1}{3},\frac{1}{3}\right)\,,
    \label{twist vectors model 1}
\end{equation}
satisfying the condition \eqref{modular conditions on twist vectors}. This model was studied in \cite{Gkountoumis:2023fym, Baykara:2023plc}, so we can use some of the results here. The massless spectrum of this orbifold consists of the $\mathcal{N}=2, D=5$ gravity multiplet and six vector multiplets, as can be read off from \autoref{huiberttable}. Notice that this is an example where accidental massless modes arise, as $m_1+m_3=2\pi$.

The choice of lattice (which can be easily checked that satisfies \eqref{modcond2}) is
\begin{equation}
    \Gamma^{5,5}=\Gamma^{4,4}(A_2\oplus A_2)\oplus\Gamma^{1,1}\ .
    \label{Narain lattice A2+A2}
\end{equation}
The sublattice $I\subset \Gamma^{4,4}(A_2\oplus A_2)$ that is invariant under the twist is
\begin{equation}
    I=\Lambda_R(A_2\oplus A_2)\,.
    \label{invarian sublattice for zero hyper model I}
\end{equation}
This can be understood by noting that the invariant lattice is spanned by the vectors $(p_L,0)$, which combined with the condition ${p}_L-{p}_R\in \Lambda_R(A_2\oplus A_2)$ yields ${p}_L\in \Lambda_R(A_2\oplus A_2)\subset \Lambda_W (A_2\oplus A_2)$. Recall that the orbifold acts as a shift on $\Gamma^{1,1}$ and leaves this lattice invariant. So, the complete orbifold invariant lattice is
\begin{equation}
    \hat{I}=\Lambda_R(A_2\oplus A_2) \oplus\Gamma^{1,1}\ .
\end{equation}

Let us now proceed with the calculation of $\mathcal{C}$. Given the twist vectors \eqref{twist vectors model 1}, we find that the equations \eqref{general commutation relations for h,g} can be solved by the matrices
\begin{equation}
 \hat{a}: \text{arbitrary}\,\qquad   \hat{b}=\begin{pmatrix}
        0_{5\times 4} & b_{5\times 1}\end{pmatrix}\,,\qquad \hat{c}= \begin{pmatrix}
            0_{4\times 5}\\
            c_{1\times 5}
        \end{pmatrix}\,,\qquad \hat{d}=\begin{pmatrix}
           d_{4\times 4} &0_{4\times 1}\\
            0_{1\times 4} &  d_{1\times 1}
        \end{pmatrix}\,,
\end{equation}
where $b_{5\times 1}$, $c_{1\times 5}$ and $d_{1\times 1}$ are arbitrary matrices, while $d_{4\times 4}$ is of the form \eqref{d SO(4) subgroup}. Once again, we employ the isomorphism \eqref{GL isomorphism} in order to bring  $d_{4\times 4}$ to the form \eqref{d4 matrix complex rep}. Now, by imposing the constraints \eqref{matrices a,b,c,d constraints} we find
    \begin{align}
       & \hat{a}_{5\times 5}^t\hat{a}_{5\times 5}-c_{5\times 1}^tc_{1\times 5}=1_{5\times 5}\,,\label{al1}\\
       & \hat{a}_{5\times 5}^tb_{5\times 1}-c_{5\times 1}^td_{1\times 1}=0_{5\times 1}\,,\label{al2}\\
       &d_{1\times 1}^td_{1\times 1}-b_{1\times 5}^tb_{5\times 1}=1_{1\times1}\,,\label{al3}\\
       &d_{4\times 4}^td_{4\times 4}=1_{4\times4}\,.\label{al4}
          \end{align}
From \eqref{al1}-\eqref{al3} it immediately follows that the matrices $\hat{a}_{5\times 5}$, $b_{5\times 1}$, $c_{1\times 5}$ and $d_{1\times 1}$ form an $\text{O}(5,1)$ group. Also, \eqref{al4} together with \eqref{d4 matrix complex rep} imply that $d_{4\times 4}\in \text{U}(2)\subset \text{O}(4)$. By imposing det$(\hat{h})=1$ we conclude that 
\begin{equation}\label{dualitygroup1}
   \mathcal{C}=\text{SO}(5,1)\times \text{SU}(2) \times \text{U}(1)\,,
\end{equation}
which, according to section \ref{OSS}, yields
\begin{equation}\label{modspace1}
    {\cal M}_V=\mathbb{R}^+\times \frac{\text{SO}(5,1)}{\text{SO}(5)}\ ,
\end{equation}
as was anticipated in \cite{Gkountoumis:2023fym}. Once again, the $\mathbb{R}^+$ factor is parametrized by the (five-dimensional) string coupling.

\subsection{A model with one hypermultiplet}
\label{z6 with 1 hyper 4 vectors}
Next we focus on models with one hypermultiplet. Surprisingly, we could find only one such model, which, besides the one hypermultiplet, also has four vector multiplets. It is a $\mathbb{Z}_6$
orbifold specified by the mass parameters
\begin{equation}
    m_1=\frac{\pi}{3}\ , \qquad m_2=\frac{2\pi}{3}\ ,\qquad m_3=\frac{\pi}{3}\ , \qquad m_4=0\ ,
\end{equation}
corresponding to the twist vectors 
\begin{equation}
    \tilde{u}=\left(\frac{1}{3},0\right)\ ,\qquad u=\left(\frac{1}{3},\frac{1}{3}\right)\ .
\end{equation}
This is a $\mathbb{Z}_3$ rotation on the bosons, but it acts as a $\mathbb{Z}_6$ on the R-vacua\footnote{From \autoref{huiberttable} we can see that orbifold charge of states in the R-sector is of the form $e^{\pm im_i}$.}. 

The massless spectrum of this model is given by $\mathcal{N}=2$ supergravity coupled to four vector multiplets and one hypermultiplet. The lattice we start with is $\Gamma^{4,4}(A_2\oplus A_2)\oplus\Gamma^{1,1}$. It can be easily verified that the twist vectors satisfy the conditions \eqref{modular conditions on twist vectors} and also
\begin{equation}
    p\mathcal{M}_{\theta}^{\,3}p\equiv p_L\mathcal{N}_L^{\,3}p_L - p_R\mathcal{N}_R^{\,3}p_R=p_L^2-p_R^2 \in 2\mathbb{Z}\ ,
\end{equation}
and so, modular invariance is ensured. The sublattice $I\subset \Gamma^{4,4}(A_2\oplus A_2)$ that is invariant under the twist is
\begin{equation}
    I=\Lambda_R(A_2)\,.
    \label{invarian sublattice for one hyper model}
\end{equation}
This can be understood by first writing $(p_L,p_R)=(p_L^1,p_L^2,p_R^1,p_R^2)$\footnote{Here the labels $1,2$ correspond to the complex torus coordinates $W^{1,2}$.} and then noting that the invariant lattice is spanned by the vectors $(0,p_L^2,0,0)$. This result combined with the condition ${p}_L-{p}_R\in \Lambda_R(A_2\oplus A_2)$, which is equivalent to ${p}_L^i-{p}_R^i\in \Lambda_R(A_2), i=1,2$, yields ${p}_L^2\in \Lambda_R(A_2)\subset \Lambda_W (A_2)$. Recall that the orbifold acts as a shift on $\Gamma^{1,1}$ and leaves this lattice invariant. So, the complete orbifold invariant lattice is
\begin{equation}
    \hat{I}=I \oplus\Gamma^{1,1}\ .
\end{equation}
Let us now focus on the calculation of $\mathcal{C}$. By comparing this model with the one presented in section \ref{Z12 example with zero hypers} we can immediately see that the matrices $\hat{a},\hat{d}$ should have exactly the same form as in \eqref{z12 matrices a,b,c,d}. For the matrices $\hat{b},\hat{c}$ we find
\begin{equation}
    \hat{b}=\begin{pmatrix}
           b^1_{2\times 2} &b^2_{2\times 2} & 0_{2\times 1}\\
            0_{3\times 2} &  0_{3\times 2}  &b_{3\times 1} \end{pmatrix}\,,\qquad\hat{c}=\begin{pmatrix}
            c^1_{2\times 2} &0_{2\times 3}\\
            c^2_{2\times 2} &0_{2\times 3}\\
            0_{1\times 2} &  c_{1\times 3}  \end{pmatrix}\,,
\end{equation}
where the matrices $ b^i_{2\times 2}, c^i_{2\times 2}, i=1,2$ are of the same form as the matrix $a^2_{2\times 2}$ in \eqref{matrix a form commuting}, while $b_{3\times 1},c_{1\times 3}$ are arbitrary. Specifically, the matrix $\hat{h}$ that commutes with the twist reads
\begin{equation}
  \hat{h}=  \begin{pmatrix}
        a & \tilde{a} & 0 &0 &0 &b_1&\tilde{b}_1&b_2&\tilde{b}_2&0\\
         -\tilde{a}&a & 0 &0 &0 &-\tilde{b}_1&b_1&-\tilde{b}_2&b_2&0\\ 
         0&0&a_1&a_2&a_3&0&0&0&0&b_3\\
         0&0&a_4&a_5&a_6&0&0&0&0&b_4\\
         0&0&a_7&a_8&a_9&0&0&0&0&b_5\\
          c_1 & \tilde{c}_1 & 0 &0 &0 &d_1&\tilde{d}_1&d_2&\tilde{d}_2&0\\
           -\tilde{c}_1&c_1 & 0 &0 &0 &-\tilde{d}_1&d_1&-\tilde{d}_2&d_2&0\\
            c_2 & \tilde{c}_2 & 0 &0 &0 &d_3&\tilde{d}_3&d_4&\tilde{d}_4&0\\
           -\tilde{c}_2&c_2 & 0 &0 &0 &-\tilde{d}_3&d_3&-\tilde{d}_4&d_4&0\\
           0&0&c_3&c_4&c_5&0&0&0&0&d_5
    \end{pmatrix}\,.
\end{equation}
By rearranging some columns and rows of $\hat{h}$ we obtain
\begin{equation}
  \hat{h}= \begin{pmatrix}
        a & b_1 & 0 &0 &0 &b_2&\tilde{a}&\tilde{b}_1&\tilde{b}_2&0\\
         c_1&d_1 & 0 &0 &0 &d_2&\tilde{c}_1&\tilde{d}_1&\tilde{d}_2&0\\ 
         0&0&a_1&a_2&a_3&0&0&0&0&b_3\\
         0&0&a_4&a_5&a_6&0&0&0&0&b_4\\
         0&0&a_7&a_8&a_9&0&0&0&0&b_5\\
          c_2 & {d}_3 & 0 &0 &0 &d_4&\tilde{c}_2&\tilde{d}_3&\tilde{d}_4&0\\
           -\tilde{a}&-\tilde{b}_1 & 0 &0 &0 &-\tilde{b}_2&a&b_1&b_2&0\\
            -\tilde{c}_1& -\tilde{d}_1 & 0 &0 &0 &-\tilde{d}_2&c_1&d_1&{d}_2&0\\
           -\tilde{c}_2&-\tilde{d}_3 & 0 &0 &0 &-\tilde{d}_4&c_2&{d}_3&d_4&0\\
           0&0&c_3&c_4&c_5&0&0&0&0&d_5
    \end{pmatrix}\,.
\end{equation}
In this form, we can see by using the isomorphism \eqref{GL isomorphism} and imposing the constraints \eqref{matrices a,b,c,d constraints} that the matrices $a^1_{2\times 2}, b^i_{2\times 2}, c^i_{2\times 2} $ (for $i=1,2$) and $d_{4\times 4}$ form a $\text{U}(1,2)$ subgroup of $\text{O}(2,4)$. Also, it follows immediately that the matrices $a_{3\times 3},b_{3\times 1},c_{1\times 3}, d_{1\times 1}$ form an $\text{O}(3,1)$ group. Finally, by demanding $\text{det}(\hat{h})=1$, we find 
 \begin{equation}
     \mathcal{C}=\text{SO}(3,1)\times \text{SU}(1,2)\times \text{U}(1)\ .
      \label{T-orbi}
 \end{equation}  
The constraints from supergravity and the number of vector multiplets and hypermultiplets, $n_V=4$ and $n_H=1$ respectively, fix the complete answer for the moduli space. It is given by
\begin{equation}
    {\cal M}_{\mathcal{O}}={\cal M}_V\times {\cal M}_H\ ,
\end{equation}
where the vector multiplet scalars live in 
\begin{equation}
    {\cal M}_V=\mathbb{R}^+\times \frac{\text{SO}(3,1)}{\text{SO}(3)}\ ,
\end{equation}
and the hypermultiplet scalars on the quaternion-K{\"a}hler manifold 
\begin{equation}
{\cal M}_H=\frac{\text{SU}(1,2)}{\text{U}(2)}\ .
\end{equation}
This space is sometimes called the universal hypermultiplet moduli space. It is important to stress that the metric on this space does not receive quantum corrections as the dilaton sits in a vector multiplet.

\subsection{Models with two hypermultiplets}

\subsubsection{$\mathbb{Z}_6$ with $n_V=2$ and $n_H=2$}
\label{z6 orbifold with nv=vh=2}

The first example of this section is the orbifold with mass parameters given by  
\begin{equation}
    m_1=\pi \ ,\qquad m_2=\frac{\pi}{3}\ , \qquad m_3=\frac{2\pi}{3}\ ,\qquad m_4=0\ .
\end{equation}
The corresponding twist vectors are
\begin{equation}
    \tilde{u}=\left(\frac{5}{6},\frac{1}{6}\right)\ ,\qquad u=\left(\frac{1}{6},\frac{1}{6}\right)\ .
    \label{z6 twist class}
\end{equation}
Once again, we choose the lattice $\Gamma^{4,4}(A_2\oplus A_2)\oplus \Gamma^{1,1}$ and we verify that the twist vectors satisfy the conditions \eqref{modular conditions on twist vectors}. Also,
\begin{equation}
    p\mathcal{M}_{\theta}^{\,3}p\equiv p_L\mathcal{N}_L^{\,3}p_L - p_R\mathcal{N}_R^{\,3}p_R=-(p_L^2-p_R^2) \in 2\mathbb{Z}\ .
\end{equation}
So, modular invariance is guaranteed. Notice that only $\Gamma^{1,1}$ is left invariant under the twist.

Regarding the massless spectrum of this model, except for the gravity multiplet one finds two vector multiplets and two hypermultiplets.  One hypermultiplet consists of accidental massless modes, because $m_1+m_2+m_3=2\pi$. The particular twist specified by \eqref{z6 twist class} falls into the class of twists studied in \cite{cardoso1994moduli}, and $\mathcal{C}$ can be easily determined using the techniques developed in the previous examples. The result is 
\begin{equation}
\mathcal{C}=\text{SO}(1,1)\times \text{SU}(2,2)\times \text{U}(1)\ .
\end{equation}
This fixes the moduli space to be
\begin{equation}
    {\cal M}_{\mathcal{O}}={\cal M}_V\times {\cal M}_H\ ,
\end{equation}
where the vector multiplet scalars live in 
\begin{equation}
    {\cal M}_V=\mathbb{R}^+\times \mathbb{R}^+\ .
\end{equation}
This space is parametrized by the string coupling and the radius of the circle and is the same as that in the example discussed in \ref{z12 with nv=2 and nh=0}. The hypermultiplet scalars form the quaternion-K{\"a}hler manifold of real dimension 8 (with $\text{SU}(2,2)\simeq \text{SO}(4,2)$),
\begin{equation}
{\cal M}_H=\frac{\text{SU}(2,2)}{\text{SU}(2)\times \text{SU}(2)\times \text{U}(1)}\simeq \frac{\text{SO}(4,2)}{\text{SO}(4)\times \text{SO}(2)}\ .
\end{equation}

\subsubsection{$\mathbb{Z}_4$ with $n_V=4$ and $n_H=2$}

As a second example, we consider the orbifold with mass parameters given by 
\begin{equation}
    m_1=\frac{\pi}{2} \ ,\qquad m_2=\pi\ ,\qquad m_3=\frac{\pi}{2}\ ,\qquad m_4=0\ .
\end{equation}
The corresponding twist vectors are
\begin{equation}
    \tilde{u}=\left(\frac{1}{2},0\right)\ ,\qquad u=\left(\frac{1}{2},\frac{1}{2}\right)\ ,
    \label{z4 twist vectors}
\end{equation}
which satisfy the conditions \eqref{modular conditions on twist vectors}. The massless spectrum of this model consists of the  gravity multiplet coupled to two hypermultiplets and four vector multiplets. One hypermultiplet is again made up by accidental massless modes, as $m_1+m_2+m_3=2\pi$. The lattice we choose is\footnote{Here we mention the isomorphism $D_2\cong A_1\oplus A_1$.} $\Gamma^{4,4}(D_2\oplus D_2)\oplus \Gamma^{1,1}$, and  we check 
\begin{equation}
    p\mathcal{M}_{\theta}^{\,2}p\equiv p_L\mathcal{N}_L^{\,2}p_L - p_R\mathcal{N}_R^{\,2}p_R=p_L^2-p_R^2 \in 2\mathbb{Z}\ .
\end{equation}
Hence, modular invariance is ensured. Following similar arguments as in section \ref{z6 with 1 hyper 4 vectors}, we find that the invariant lattice is
\begin{equation}
    \hat{I} = \Lambda_R(D_2)\oplus \Gamma^{1,1}\,.
\end{equation}

Given the simple form of the twist vectors \eqref{z4 twist vectors} and the analysis of the previous sections, it is easy to see that $\hat{h}$ takes the form
\begin{equation}
    \hat{a}=\begin{pmatrix}
           a_{2\times 2} &0_{2\times 3}\\
            0_{3\times 2} &  a_{3\times 3}
        \end{pmatrix}\,,\quad  \hat{b}=\begin{pmatrix}
           b^1_{2\times 2} &b^2_{2\times 2} & 0_{2\times 1}\\
            0_{3\times 2} &  0_{3\times 2}  &b_{3\times 1} \end{pmatrix}\,,\quad\hat{c}=\begin{pmatrix}
            c^1_{2\times 2} &0_{2\times 3}\\
            c^2_{2\times 2} &0_{2\times 3}\\
            0_{1\times 2} &  c_{1\times 3}  \end{pmatrix}\,,\quad  \hat{d}=\begin{pmatrix}
           d_{4\times 4} &0_{4\times 1}\\
            0_{1\times 4} &  d_{1\times 1}
        \end{pmatrix}\,.
\end{equation}
Imposing the constraints \eqref{matrices a,b,c,d constraints} and $\text{det}(\hat{h})=1$ yields
\begin{equation}
  \mathcal{C}=\text{S}\left[\text{O}(3,1)\times \text{O}(2,4)\right]\ .
\end{equation}
The vector multiplet and hypermultiplet moduli spaces are
\begin{equation}
    {\cal M}_V=\mathbb{R}^+\times \frac{\text{SO}(3,1)}{\text{SO}(3)}\ ,\qquad {\cal M}_H=\frac{\text{SO}(4,2)}{\text{SO}(4)\times \text{SO}(2)}\ .
   \end{equation}

\subsection{Moduli spaces and $\mathcal{N}=2, D=5$ supergravity}

Here we make some general remarks about the moduli spaces that we found. All models presented in this section give vector multiplet moduli spaces that  are of the general form 
\begin{equation}\label{VMMod}
    {\cal M}_V=\mathbb{R}^+\times \frac{\text{SO}(n_V-1,1)}{\text{SO}(n_V-1)}\ ,
\end{equation}
consistent with the constraints from real special geometry in $\mathcal{N}=2,D=5$, with $n_V$ the number of vector multiplets. Indeed, when the moduli space factorises, it must be of the form \eqref{VMMod} 
\cite{Gunaydin:1983bi,Gunaydin:1983rk}.
The cubic polynomial of real special geometry is then of the factorised form 
\begin{equation}
    C(h)=h^1 Q(h^I)\ ,\qquad Q(h^I)=h^I\bar{\eta}_{IJ}h^J\ ,
\end{equation}
with $I=(i,2)\,,i=3,...,n_V+1$ and where $\bar{\eta}$ is a quadratic form of (mostly minus) signature $(n_V-1,1)$, making the SO$(n_V-1,1)$ isometry manifest. The factorization of $\mathcal{M}_V$ is a special property and not the most general $\mathcal{N}=2, D=5$ supergravity coupling of vector multiplets. It is a consequence of the underlying $\mathcal{N}=8$ supersymmetry which is spontaneously broken, and the fact that $\text{E}_{6(6)}$ has a maximal subgroup which factorises as $\text{SO}(1,1) \times \text{Spin}(5,5)$, as explained in section \ref{OSS}. 

Hypermultiplet scalars coupled to supergravity with eight supercharges parametrize (non-compact) quaterion-K\"ahler (QK) manifolds. The cases arising in this paper correspond to homogeneous spaces $G/H$. These are classified, and the ones appearing in this section belong to the series
\begin{equation}
    X(n_H)=\frac{\text{SU}(n_H,2)}{\text{SU}(n_H)\times \text{U}(2)} \ , \qquad Y(n_H)=\frac{\text{SO}(n_H,4)}{\text{SO}(n_H)\times \text{SO}(4)}\ ,
\end{equation}
with $n_H$ the number of hypermultiplets. The examples we discussed in this section correspond to $X(1)$ and $X(2)$. Note that $X(2)\simeq Y(2)$ up to double covers that for the purpose of this paper are not relevant and therefore not explicitly written. 

This concludes the discussion of the examples. In the next section, we turn to their dual versions.

\section{Dual pairs}
\label{dual pairs}
In this section we construct dual pairs \`a la Sen and Vafa \cite{sen1995dual}. We will briefly discuss the procedure of constructing dual pairs here; for more details we refer to \cite{sen1995dual}. The starting point is type II string theory compactified on $T^4$, with moduli space
$\mathcal{M}_{T^4}=\text{Spin}(5,5)/\left[\text{Spin}(5)\times \text{Spin}(5)\right]$.
The U-duality group of the theory is $\text{Spin}(5,5;\mathbb{Z})$ which contains the T-duality group  $\text{Spin}(4,4;\mathbb{Z})$. Take two elements $g,g'\in \text{Spin}(4,4;\mathbb{Z})$ of order $p$ that are conjugate in $\text{Spin}(5,5;\mathbb{Z})$ but not in $\text{Spin}(4,4;\mathbb{Z})$, that is
\begin{equation}
    \hat{g} \,g\,\hat{g}^{-1} = g'\,,\quad \hat{g}\,\in \text{Spin}(5,5;\mathbb{Z})\,.
    \label{general duality element}
\end{equation}
Consider a point in the moduli space $m\in \mathcal{M}_{T^4}$
that is invariant under the action of $g$, so that $g \cdot m=m$. Then the point $m'=\hat g \cdot m\in \mathcal{M}_{T^4}$ is invariant under the action of $g'$. 

Next, compactify on a  further circle $S^1$. At the point in the moduli space $m\in \mathcal{M}_{T^4}$
we can orbifold by the action of $g$ combined with a shift by $2\pi \mathcal{R}/p$ on the circle and at
the point in the moduli space $m'\in \mathcal{M}_{T^4}$
we can orbifold by the action of $g'$ combined with a shift by $2\pi \mathcal{R}'/p$ on the  circle.
The two orbifold theories should then be equivalent non-perturbatively
\cite{Vafa:1995gm,sen1995dual}, with a relation between $\mathcal{R}$ and $\mathcal{R}'$ that we specify below.

In the context of type IIB string theory, the particular element $\hat{g}$ chosen in \cite{sen1995dual} can be expressed in terms of the SL$(2;\mathbb{Z}$) S-duality element $s$ that inverts the ten-dimensional axion-dilaton field, and the SO$(4,4;\mathbb{Z}$) T-duality element $t_{67}\cdot t_{89}$, 
where $t_{ij}$ inverts the
 K{\"a}hler modulus of the $i-j$ plane, so that
\begin{equation}
    \hat{g} = s\cdot t_{67}\cdot t_{89} \cdot s^{-1}\,.
    \label{duality element omega}
\end{equation}
If the R-R scalars are set to zero, this $\hat{g}$ acts on the six-dimensional dilaton $\phi_6$ and the NS-NS three-form $H_3$ exactly as string-string duality acts on these fields in six dimensions \cite{sen1995dual}, that is (see also \cite{Duff:1994zt,Witten:1995ex})
\begin{equation}
    \phi_6'=-\phi_6\,,\qquad H_3' = e^{-2\phi_6}*H_3\,.
\end{equation}This induces the following transformation on the 6-dimensional metric in string frame (in Einstein frame, the metric is invariant)
\begin{equation}
    G_{\hat{\mu}\hat{\nu}}'= e^{-2\phi_6}G_{\hat{\mu}\hat{\nu}}\,, \qquad \hat{\mu},\hat{\nu}=0,\ldots,5\,.
\end{equation}
Also, it follows that under string-string duality the strong coupling regime of one theory corresponds to the weak coupling regime of the dual theory
as 
\begin{equation}
   \phi_6'=-\phi_6\implies \lambda_6'=\frac{1}{\lambda _6}
\end{equation}
 where $\lambda _6=\langle e^{\phi_6}\rangle$ and similarly for $\lambda_6'$.
Moreover, we can determine how the duality works upon further compactification to five dimensions \cite{Witten:1995ex}; this will be essential for our discussion later in this section. We have\footnote{For convenience, we set $\alpha '=1$ from now on.}
\begin{equation}
    G_{55}'= e^{-2\phi_6}G_{55}\implies\mathcal{R}'=\frac{\mathcal{R}}{\lambda_6}\,.
\end{equation}
We can rewrite the above equations in terms of the $5D$ string coupling constant $\lambda_5$, using $\lambda_6=\sqrt{\mathcal{R}}\lambda_5$ and similarly for $\lambda_6'$. We obtain
\begin{equation}
    \lambda_5'=\frac{1}{\sqrt{\lambda_5}\mathcal{R}^{{3}/{4}}}\,,\qquad \mathcal{R}'=\frac{\sqrt{\mathcal{R}}}{\lambda_5}\,.
    \label{5d string-string duality}
\end{equation}
We can also perform a T-duality transformation 
taking the circle of radius $\mathcal{R}'$ to the T-dual circle of radius $\widetilde{\mathcal{R}}=1/\mathcal{R}'
$
under which the five-dimensional string coupling remains invariant, $ \lambda_5'=\widetilde{\lambda}_5$,
so that $\lambda'_5=\lambda'_6/\sqrt{\mathcal{R}'}
=\widetilde\lambda_6/\sqrt{\widetilde{\mathcal{R}}}$ 
implies
\begin{equation}
   \lambda_6'=\frac{\widetilde{\lambda}_6}{\widetilde{\mathcal{R}}}\,.
\end{equation}
After the T-duality transformation, \eqref{5d string-string duality} becomes
\begin{equation}
     \widetilde{\lambda}_5=\frac{1}{\sqrt{\lambda_5}\mathcal{R}^{{3}/{4}}}\,,\qquad \widetilde{\mathcal{R}}=\frac{\lambda_5}{\sqrt{\mathcal{R}}}\,.
     \label{T-dual string-string duality}
\end{equation}
Finally, we can rewrite \eqref{T-dual string-string duality} in terms of the six-dimensional couplings as
\begin{equation}
     \widetilde{\lambda}_6=\frac{1}{\mathcal{R}}\,,\qquad \widetilde{\mathcal{R}}=\frac{\lambda_6}{{\mathcal{R}}}\,.
\end{equation}
For the purposes of this work, we start from type IIB theory and, following the procedure presented above (cf.\,\eqref{general duality element}-\eqref{duality element omega}), we construct a dual pair of type IIB orbifolds, $\mathcal{O}_{\text{IIB}}$ and $\mathcal{O}_{\text{IIB}}'$. Then we apply a T-duality transformation on $\mathcal{O}_{\text{IIB}}'$ and we obtain a T-dual type IIA orbifold $\widetilde{\mathcal{O}}_{\text{IIA}}$. Note that the duality between $\mathcal{O}_{\text{IIB}}$ and $\mathcal{O}_{\text{IIB}}'$ is non-perturbative, while the duality between $\mathcal{O}_{\text{IIB}}'$ and $\widetilde{\mathcal{O}}_{\text{IIA}}$ is perturbative, and the models $\mathcal{O}_{\text{IIB}}'$ and $\widetilde{\mathcal{O}}_{\text{IIA}}$ have the same perturbative spectrum and moduli space. Now, according to \eqref{T-dual string-string duality}, the strong coupling regime of the $\mathcal{O}_{\text{IIB}}$ model at fixed radius corresponds to weak coupling and large radius of the dual $\widetilde{\mathcal{O}}_{\text{IIA}}$ model.

Consider the $\mathbb{R}^+\times \mathbb{R}^+$ subspace of the moduli space parameterised by $\lambda_5,\mathcal{R}$.
It can equivalently be viewed as $\mathbb{R}^2$ 
 with coordinates $\phi_5,\sigma$ where
$\lambda_5=\langle e^{\phi_5}\rangle$ and $\mathcal{R}=e^{2\sigma/{\sqrt{3}}}$.
In the examples we discuss in this paper, the classical moduli space metric for this subspace is 
\begin{equation}\label{isom-metrica}
    {\rm d}s^2=
    {\rm d}\phi_5^2+{\rm d}\sigma^2\ .
\end{equation}
The U-duality
between $\mathcal{O}_{\text{IIB}}'$ and $\mathcal{O}_{\text{IIB}}$
then gives
\begin{equation}\label{dictionaryOprimeO}
    \phi_5'=-\frac{1}{2}\phi_5-\frac{\sqrt{3}}{2}\sigma\ ,\qquad \sigma'=-\frac{\sqrt{3}}{2}\phi_5+\frac{1}{2}\sigma\ ,
\end{equation}
while the T-duality
between $\mathcal{O}_{\text{IIB}}'$ and $\widetilde{\mathcal{O}}_{\text{IIA}}$ gives
\begin{equation}\label{dictionaryOtildeO}
    \tilde{\phi}_5=-\frac{1}{2}\phi_5-\frac{\sqrt{3}}{2}\sigma\ ,\qquad \tilde{\sigma}=\frac{\sqrt{3}}{2}\phi_5-\frac{1}{2}\sigma\ ,
\end{equation}
Notice that this is a rotation through an angle $2\pi/3$ and, since it is a rotation, it is an isometry of the flat metric (\ref{isom-metrica}), taking it to
\begin{equation}\label{isom-metric}
  {\rm d}s^2=  {\rm d}\tilde{\phi}_5^2+{\rm d}\tilde{\sigma}^2\ .
\end{equation}
This fact will be important in our arguments showing the absence of quantum corrections to the moduli spaces. For instance, in the two-moduli example of section \ref{z12 with nv=2 and nh=0}, the classical moduli space
 $\mathbb{R}^+\times \mathbb{R}^+$ has a
metric that is precisely the flat metric (\ref{isom-metrica}), and hence is unchanged by the strong-weak duality. We will return to this in the examples below and in the next section.

Finally, we mention here that the five-dimensional duality proposed in \cite{Witten:1995ex} was between type IIB string theory on $\mathbb{R}^{1,4}\times S^1 \times {K}3$ and heterotic string theory on $\mathbb{R}^{1,4}\times T^5$, and it was argued there that the strong coupling regime of the heterotic string compactification corresponds to the type IIB compactification at large radius. This duality was derived from string-string duality in six dimensions, which relates the heterotic string on $\mathbb{R}^{1,5}\times T^4$ with the type IIA string on $\mathbb{R}^{1,5}\times {K}3$. 

\subsection{Examples}

In this section we present explicit examples of dual type II pairs with $\mathcal{N}=2$ supersymmetry in $5D$. In practice, for the construction of a dual pair we start with an orbifold twist $\mathcal{M}_{\theta}$ specified by the twist vectors $\tilde{u},u$, that has the form
\begin{equation}
    \mathcal{M}_{\theta}= \text{diag}\left(R(2\pi\tilde{u}_3),R(2\pi\tilde{u}_4),R(2\pi u_3),R(2\pi u_4)\right)\,,
\end{equation}
according to \eqref{orbifold twist in terms of twist vectors}. We will refer to the orbifold specified by $\mathcal{M}_{\theta}$ as the \say{initial} model. Then, the action of $\hat g$ on the twist vectors is simply given by (see \cite{sen1995dual} for details)
\begin{equation}
    \begin{pmatrix}
        \tilde{u}_3'\\
        \tilde{u}_4'\\
        u_3'\\
        u_4'
    \end{pmatrix} =\frac{1}{2} \begin{pmatrix}
        1&-1&1&-1\\
        -1&1&1&-1\\
        1&1&1&1\\
        -1&-1&1&1
    \end{pmatrix}\,\begin{pmatrix}
        \tilde{u}_3\\
        \tilde{u}_4\\
        u_3\\
        u_4
    \end{pmatrix}\,.
    \label{matrix for dual pairs}
\end{equation}
This induces a duality transformation on the mass parameters, and using \eqref{relation twist vectors-mass parameters} we find
\begin{equation}
    m_1'=m_4\ ,\qquad m_2'=m_2\ ,\qquad m_3'=m_3\ ,\qquad m_4'=m_1\ ,
\end{equation}
so it simply amounts to an exchange $m_1\leftrightarrow m_4$. 
For example, starting from a twist with $m_4=0$
\begin{equation}
    \mathcal{M}_{\theta} = \text{diag}\left(R(m_1+m_3),R(m_1-m_3),R(m_2),R(m_2)\right)\,,
\end{equation}
the twist of the dual model is 
\begin{equation}
    \mathcal{M}_{\theta}' = \text{diag}\left(R(m_3),R(-m_3),R(m_2+m_1),R(m_2-m_1)\right)\,.
\end{equation}
We mention here that it is possible to start with a model for which the only massless fields are in the NS-NS sector and under the duality transformation end up with a model with both NS-NS and R-R massless fields.

\subsubsection{Dual pairs without R-R fields}
\label{dual pairs without R-R fields}
Here, we work out two concrete examples of $\mathcal{N}=2$ dual pairs in $5D$ where the massless spectra of both the initial and the dual theory consist purely of NS-NS states. This has the advantage that the moduli space of the dual theory can be found using the same techniques as for the examples of section \ref{N=2 models in 5d with exact moduli spaces}, namely by finding the commutant with the T-duality group. We could find only two such models.

\subsection*{Dual pair I:}
\label{dual pair I}
We start from the $\mathbb{Z}_{12}$ orbifold presented in section \ref{z12 with nv=2 and nh=0} with $n_V=2$ and $n_H=0$. The mass parameters and twist vectors of this model are 
\begin{equation}
     m_1=\frac{2\pi}{3}\ , \qquad m_2= \frac{\pi}{2}\ ,\qquad m_3=\frac{\pi}{3} \ ,\qquad m_4=0\,,
\end{equation}
\begin{equation}
     \tilde{u}=\left(\frac{1}{2},\frac{1}{6}\right)\ ,\qquad u=\left(\frac{1}{4},\frac{1}{4}\right)\,,
\end{equation}
and its moduli space was found to be
\begin{equation}
     {\cal M}_{\mathcal{O}}=\mathbb{R}^+\times \mathbb{R}^+\ .
     \label{moduli space dual pair i}
\end{equation}
The two scalars parametrizing this moduli space are the string coupling $\lambda_5=\langle e^{\varphi}\rangle$ and the radius of the circle $\mathcal{R}
=e^{2\sigma/\sqrt{3}}$ (recall that we are setting $\alpha'=1$), and the classical metric on ${\cal M}_{\mathcal{O}}$ is the flat metric \eqref{classmetric}.

Regarding the dual model, we can calculate the mass parameters and twist vectors using \eqref{matrix for dual pairs} and \eqref{relation twist vectors-mass parameters}. We find
\begin{equation}
     m_1'=0\ , \qquad m_2'= \frac{\pi}{2}\ ,\qquad m_3'=\frac{\pi}{3} \ ,\qquad m_4'=\frac{2\pi}{3}\,,
\end{equation}
\begin{equation}
     \tilde{u}'=\left(\frac{1}{6},-\frac{1}{6}\right)\ ,\qquad u'=\left(\frac{7}{12},-\frac{1}{12}\right)\,.
\end{equation}
This is again a $\mathbb{Z}_{12}$ orbifold with $n_V=2$ and $n_H=0$, which we denote by $\mathcal{O}'$, and all the massless fields come from the NS-NS sector. So, we can determine the moduli space $\mathcal{M}_{\mathcal{O}'}$ of this dual model using the techniques developed in sections \ref{OSS} and \ref{N=2 models in 5d with exact moduli spaces}. We find
\begin{equation}
     {\cal M}_{\mathcal{O}'}=\mathbb{R}^+\times \mathbb{R}^+\ .
     \label{moduli space dual pair i dual}
\end{equation}
Classically, in the dual variables, the metric is again the flat metric. The moduli space is parametrized by the string coupling $\lambda_5'=\langle e^{\varphi'}\rangle$ and the radius of the circle $\mathcal{R}'
=e^{2\sigma'/\sqrt{3}}$. 
T-duality takes the circle radius to  $\widetilde{\mathcal{R}}=1/\mathcal{R}'
$
 and gives the T-dual IIA orbifold model, denoted by $\widetilde{\mathcal{O}}$. Since it is obtained by a T-duality of the $\mathcal{O}'$ orbifold, we have that $\tilde\sigma=-\sigma'$ and so the moduli space is again given by
\begin{equation}
{\cal{M}}_{\widetilde{\mathcal{O}}}=\mathbb{R}^+\times \mathbb{R}^+\ ,
\end{equation}
parameterised by $\widetilde{\mathcal{R}}$ and $\widetilde{\lambda}_5$. The classical metric is again the flat one, as it is obtained from the T-dual of the $\mathcal{O}'$ theory. 

We  now consider quantum corrections to the metric on the moduli space \eqref{moduli space dual pair i}.
Consider first the initial orbifold ${\mathcal{O}}$.
For the classical theory, the moduli space metric is (\ref{isom-metrica}) and is independent of the dilaton $\phi_5$.
Quantum effects can in principle deform this moduli space metric and lead to dilaton-dependence.
Now, as we can see from \eqref{T-dual string-string duality}, 
the theory $ {\cal{O}}$ at  strong coupling (large $\lambda_5$) and fixed radius $\mathcal{R}$ is given by the dual theory $\widetilde{\cal{O}}$ at
weak coupling (in $\tilde{\lambda}_5$). However, we have seen that the classical moduli space metric for the dual
theory $\widetilde{\cal{O}}$ is again the flat metric
(\ref{isom-metric}). The classical theory $\widetilde{\cal{O}}$ is the strong-coupling limit of the original theory $ {\cal{O}}$, so we see that the strong-coupling limit of the moduli space metric on $ {\cal{O}}$
is the same as for the classical theory, greatly limiting the form any quantum corrections to the metric can take.
In the next section, we present a stronger version of this argument in which we parameterise the possible quantum corrections and then  use duality to argue that no quantum corrections in fact arise.

\subsection*{Dual pair II:}
\label{dual pair II}
Here we present another example where the massless spectra of both the initial and the dual model consist purely of NS-NS fields. The initial model is specified by the following mass parameters and twist vectors
\begin{equation}
     m_1=\pi\ , \qquad m_2= \frac{\pi}{3}\ ,\qquad m_3=\frac{2\pi}{3} \ ,\qquad m_4=0\,,
\end{equation}
\begin{equation}
     \tilde{u}=\left(\frac{5}{6},\frac{1}{6}\right)\ ,\qquad u=\left(\frac{1}{6},\frac{1}{6}\right)\,,
\end{equation}
This is a $\mathbb{Z}_6$ orbifold with $n_V=n_H=2$ discussed in section \ref{z6 orbifold with nv=vh=2}. The moduli space of this model is
\begin{equation}
{\cal M}_{\mathcal{O}}={\cal M}_V\times{\cal M}_H=\mathbb{R}^+\times \mathbb{R}^+\times \frac{\text{SO}(4,2)}{\text{SO}(4)\times \text{SO}(2)}\ ,
\label{moduli space dual pair II}
\end{equation}
where ${\cal M}_V=\mathbb{R}^+\times \mathbb{R}^+$ is parameterised by the string coupling and the radius of the circle. For the dual model we find (cf.\,\eqref{matrix for dual pairs},\eqref{relation twist vectors-mass parameters})
\begin{equation}
     m_1'=0\ , \qquad m_2'= \frac{\pi}{3}\ ,\qquad m_3'=\frac{2\pi}{3} \ ,\qquad m_4'=\pi\,,
\end{equation}
\begin{equation}
     \tilde{u}'=\left(\frac{1}{3},-\frac{1}{3}\right)\ ,\qquad u'=\left(\frac{2}{3},-\frac{1}{3}\right)\,.
\end{equation}
This is a $\mathbb{Z}_6$ orbifold with $n_V=n_H=2$, with only NS-NS massless fields. Once again, we can specify classically $\mathcal{M}_{\mathcal{O}}'$ using the methods of sections \ref{OSS} and \ref{N=2 models in 5d with exact moduli spaces}. We find that 
\begin{equation}
  \mathcal{M}_{\mathcal{O}}' = \mathbb{R}^+\times \mathbb{R}^+\times \frac{\text{SO}(4,2)}{\text{SO}(4)\times \text{SO}(2)}\,.
  \label{dual pair II dual moduli space}
\end{equation}
In section \ref{absence} we 
 argue that the classical moduli space metric on \eqref{moduli space dual pair II} is exact and receives no quantum corrections.

\subsubsection{Dual pairs with R-R fields}
\label{dual models with R-R fields}
As we mentioned in the beginning of this section, it is possible to start with a model without massless R-R fields and, after applying the transformation \eqref{matrix for dual pairs}, obtain a dual model with massless R-R fields. We collect such dual pairs in \autoref{Table of dual pairs with R-R field}. The moduli space of the initial models was specified in section \ref{N=2 models in 5d with exact moduli spaces}. Regarding the moduli space of the dual models, we could not use the techniques of sections \ref{OSS} and \ref{N=2 models in 5d with exact moduli spaces} because the spectra of those models contain R-R massless states. However, classically the moduli spaces of the dual models (with R-R massless scalars) can be determined by the constraints of spontaneously broken $\mathcal{N}=8 \to \mathcal{N}=2, D=5$ supergravity, and the classical vector multiplet moduli spaces are again of the form \eqref{VMMod}. In the next section we will present evidence for the absence of quantum corrections to the vector multiplet moduli spaces of these dual pairs as well.

\begin{table}[ht!]
\centering
{\renewcommand*{\arraystretch}{2.3}
\footnotesize{
	\begin{tabular}{|c|c|c|c|c|c|c|}
 \hline
  $\mathbb{Z}_p$	&  $(m_1,m_2,m_3,m_4)$ & $\tilde{u},u$ & $(m_1',m_2',m_3',m_4')$ & $\tilde{u}',u'$  & $\phi_{(NS,R)},A^{\mu}_{(NS,R)} $ & $(n_V,n_H)$ \\ \hline
 $\mathbb{Z}_{12}$ & $(\tfrac{2\pi}{3},\tfrac{\pi}{2},\tfrac{2\pi}{3},0)$ & $(\tfrac{2}{3},0),(\tfrac{1}{4},\tfrac{1}{4})$ & $(0,\tfrac{\pi}{2},\tfrac{2\pi}{3},\tfrac{2\pi}{3})$ & $(\tfrac{1}{3},-\tfrac{1}{3}),(\tfrac{7}{12},-\tfrac{1}{12})$ & $(2,2), (3,2)$  & $(4,0)$ \\ \hline
   $\mathbb{Z}_6$ & $(\pi,\tfrac{2\pi}{3},\pi,0)$ & $(1,0),(\tfrac{1}{3},\tfrac{1}{3})$ & $(0,\tfrac{2\pi}{3},\pi,\pi)$ & $(\tfrac{1}{2},-\tfrac{1}{2}),(\tfrac{5}{6},-\tfrac{1}{6})$  & $(2,4),(3,4)$& $(6,0)$\\ \hline
 $\mathbb{Z}_6$ & $(\tfrac{\pi}{3},\tfrac{2\pi}{3},\tfrac{\pi}{3},0)$ & $(\tfrac{1}{3},0),(\tfrac{1}{3},\tfrac{1}{3})$ &  $(0,\tfrac{2\pi}{3},\tfrac{\pi}{3},\tfrac{\pi}{3})$  &  $(\tfrac{1}{6},-\tfrac{1}{6}),(\tfrac{1}{2},\tfrac{1}{6})$ & $(6,2), (3,2)$ &$(4,1)$\\ \hline 
  ${\mathbb{Z}_4}$ & $(\tfrac{\pi}{2},\pi,\tfrac{\pi}{2},0)$ & $(\tfrac{1}{2},0),(\tfrac{1}{2},\tfrac{1}{2})$ & $(0,\pi,\tfrac{\pi}{2},\tfrac{\pi}{2})$ & $(\tfrac{1}{4},-\tfrac{1}{4}),(\tfrac{3}{4},\tfrac{1}{4})$ & $(10,2),(3,2)$ & $(4,2)$ \\ \hline
     	\end{tabular}}}
        \caption{\textit{Table of $\mathcal{N}=2$ dual pairs in $5D$. First, we write down the mass parameters and twist vectors of the initial models, which have no R-R massless states. Then we write down the mass parameters and the twist vectors of the dual models, which do have R-R massless states, and we specify the number of their NS-NS (NS) and R-R (R) scalars ($\phi$) and vectors ($A^{\mu}$). In the last column we write down the number of vector multiplets $n_V$ and hypermultiplets $n_H$, which is of course the same both for the initial model and its dual.}}
        \label{Table of dual pairs with R-R field}
\end{table}

\section{Absence of quantum corrections to the vector multiplet moduli space}\label{absence}

In this section, we argue
that there are no quantum corrections to the vector multiplet moduli space of the models we have constructed, using the properties of dual pairs discussed in the previous section. This furthermore leads to predictions about the absence of certain Chern-Simons couplings in the effective action. 

\subsection{Dual NS-NS pairs with $n_V=2$}

In this subsection we discuss  the possibility of quantum corrections for the two examples presented in sections \ref{z12 with nv=2 and nh=0} and \ref{z6 orbifold with nv=vh=2}. These two examples only differ in the number of hypermultiplets, but it is already known that the hypermultiplet moduli space metric does not receive quantum corrections. Consequently, we  focus on the vector multiplet moduli space metric. For both models, the classical vector multiplet moduli space metric is given by the flat metric (\ref{isom-metrica}) on $\mathbb{R}^2$. In $\mathcal{N}=2, D=5$ supergravity, the vector multiplet moduli space metric is completely characterised by the $d$-symbols, which are real numbers that do not depend on moduli such as the string coupling.
However, as we shall see below, it is in principle possible that the 
$d$-symbols for the quantum theory can jump and are different from those of the classical theory. Such changes to the $d$-symbols can be computed from loop effects that induce transitions in the Chern-Simons couplings, see e.g.\ \cite{witten:1996qb}.

We first notice that the classical values for the $d$-symbols \eqref{d-symbols1} can also be determined from the Scherk-Schwarz reduction from six to five dimensions. Indeed, as was shown in \cite{Hull:2020byc}, the $d$-symbols \eqref{d-symbols1} follow from the reduction of the $6D$ self-dual and anti-self-dual tensors that remain massless under the twist. It was shown explicitly in \cite{Hull:2020byc} that they generate the $5D$ Chern-Simons couplings $d_{ABC}A^A\wedge F^B \wedge F^C$, with $d_{122}=-d_{133}=1$, where $A^{1}$ corresponds to the graviphoton, i.e.\ the Kaluza-Klein vector coming from the metric. Furthermore, classically, $d_{111}=0$, since no Chern-Simons term appears with only the graviphoton. At the quantum level, however, $d_{111}$ can be generated from  one-loop effects by integrating out charged matter, as was also discussed  in \cite{Hull:2020byc} (see e.g.\ \cite{Antoniadis:1995vz,Bonetti:2013cza} for some further literature on loop corrections to Chern-Simons terms). 

The geometry of the scalar manifold could then be corrected by a term proportional to $d_{111}$. We now claim that the only possible quantum correction to the cubic polynomial is
\begin{equation}\label{pertC1}
    C(h)=d_{ABC}h^Ah^Bh^C=3h^1\Big((h^2)^2-(h^3)^2\Big)+d_{111}(h^1)^3\, .
\end{equation}
This
consists of the classical term proportional to $h^1$ and   perturbative corrections proportional to $(h^1)^3$.
Such a term, proportional to $(h^1)^3$, cannot be absorbed into a field redefinition, so it would be a genuine deformation. 
Notice that it furthermore breaks the continuous SO(1,1) dilatonic scale transformation but not its quantum discrete subgroup $\mathbb{Z}_2$, which acts trivially on the dilaton. 

There is another argument why the $(h^1)^3$ term is the only possible quantum correction. Any other cubic term can not be added, since it is either of the same order in $h^1$, i.e.\ a classical term, or it is forbidden by the quantum symmetry $\mathbb{Z}_2$, which sends $h^{2,3}\to -h^{2,3}$ (see the discussion in the end of section \ref{z12 with nv=2 and nh=0}). In particular, a term quadratic in $h^1$ should be accompanied by linear terms in $h^{2,3}$ but these are forbidden by the aforementioned $\mathbb{Z}_2$. This is consistent with the fact that at one-loop, only Chern-Simons terms with external graviphotons can be generated. It is furthermore also consistent with the fact that there are no $\alpha'$ corrections, as the radius sits in the variables $h^2$ and $h^3$ and any correction in $\alpha'$ must come in the combination $\mathcal{R}/\sqrt{\alpha'}$.

The moduli space metric following from (\ref{pertC1}) is 
found as follows. In order to solve $C(h)=1$ in \eqref{pertC1}, we parametrize $h^2$ and $h^3$ as in \eqref{repsh}, and solve for $h^1$ perturbatively in powers of the string coupling, 
\begin{equation}
\label{changex}
h^1=\frac{1}{3a^2}e^{2\varphi/3}\Big(1-\frac{d_{111}}{27a^6}\,e^{2\varphi}+\cdots\Big)\ .
\end{equation}
Plugging this back into the metric ${\rm d}s^2=-9\,d_{ABC}h^A{\rm d}h^B{\rm d}h^C$ gives the perturbative quantum corrections to the metric. 
With the identification $\phi_5=\varphi$, this gives a power series in $e^{2\phi_5}$.
The resulting metric is then of the form
\begin{equation}\label{isom-metricb}
    {\rm d}s^2=
    \left( \delta_{ab}+  h_{ab} (e^{2\phi_5})\right)
    {\rm d}x^a {\rm d}x^b\ ,
\end{equation}
where $x^a=(\phi_5,\sigma)$. This is the classical metric (\ref{isom-metrica}) plus a 
correction $h_{ab} (e^{2\phi_5})$ comprising the quantum corrections which is a power series  in $e^{2\phi_5}$.
In this case, there are no non-perturbative contributions. (We could in principle at this stage consider the possibility of non-perturbative contributions to $h_{ab} (e^{2\phi_5})$. Our argument below would then confirm that  there are no non-perturbative contributions.)

In section \ref{dual pair I} we constructed the dual model, and we found  that the classical vector multiplet moduli space metric of the dual theory is again  flat. The further T-duality discussed in section \ref{dual pair I} then gives another dual model with a flat moduli space metric (\ref{isom-metric}). On the other hand, the metric for the moduli space of the dual theory can  be obtained from (\ref{isom-metricb}) by
the coordinate transformation (\ref{dictionaryOtildeO}).
The classical limit of the resulting dual metric is obtained by setting $\tilde{\phi}_5=0$ and the components of this metric have non-trivial dependence on
$\tilde{\sigma}$ if $h_{ab}\ne 0$. 
Explicitly, transforming to the tilde coordinates and setting $\tilde{\phi}_5=0$ amounts to transforming $\phi_5$ to $\sqrt 3 \tilde\sigma$. For example, if $h_{ab}$ is independent of $\sigma$,
then the classical metric of the dual moduli space is
\begin{equation}\label{isom-metricdual}
    {\rm d}s^2=
    \left( \delta_{ab}+ 
   \tilde  h_{ab} (e^{2\sqrt 3 \tilde\sigma})
     \right)
    {\rm d}\tilde x^a {\rm d}\tilde x^b\ .
\end{equation}
(If $h_{ab}$ depends explicitly on $\sigma$, then there would be a similar but more complicated formula.)
However,  the classical moduli space metric for the dual theory is (\ref{isom-metric}) and so   these can only agree if $h_{ab}= 0$.
Therefore, we conclude that the classical moduli space is exact, and no quantum corrections arise.

The absence of quantum corrections leads to the prediction that in the quantum theory, $d_{111}$ vanishes. This means that the result of integrating out all $A^1$-charged states in string theory cancels out. This is a non-trivial prediction, which we comment further about in the discussion.

\subsection{Dual pairs with R-R fields}

The classical values for the $d$-symbols in all the examples discussed in this paper are of the form
\begin{equation}\label{dsymb}
d_{122}=1\ ,\qquad d_{1ij}=-\delta_{ij}\ ,\qquad i=3,...,n_{V+1}\ ,
\end{equation}
and correspond to factorizable moduli spaces \eqref{VMMod}. They are homogeneous spaces and fall in the classification of \cite{deWit:1991nm,deWit:1992wf}\footnote{\label{basisL(0,P)}The parametrization of the $d$-symbols given in \eqref{dsymb} is the one from \cite{Gunaydin:1983rk}. It is related to the parametrization given in \cite{deWit:1991nm} by a basis transformation. The moduli spaces \eqref{VMMod} are denoted by $L(0,P)$ in \cite{deWit:1991nm,deWit:1992wf} and are a special case of the spaces $L(0,P,\dot P)$ with $\dot P=0$. They are all homogeneous spaces and for $\dot P=0$ become symmetric spaces that furthermore factorise as in \eqref{VMMod}, see equation (4.10) in \cite{deWit:1991nm}.}. The isometry group $\text{SO}(n_V-1,1)\subset \mathcal{K}$ acts linearly on $h^I=(h^i,h^2)$. In the quantum theory, this group is contained in the U-duality group and should be taken over the integers. For an even number of vector multiplets, it contains the element $-1_{n_V\times n_V}$ which sends $h^I\to - h^I$. The vector in the direction of $h^1$ corresponds to the graviphoton, and classically, $d_{111}=0$ (recall that no Chern-Simons term appears with only the graviphoton). As in the previous examples, at the quantum level  $d_{111}$ could be generated from a one-loop induced Chern-Simons term.
The metric on the moduli space could then be corrected with  terms proportional to $d_{111}$. The only possible quantum correction to the cubic polynomial is
\begin{equation}
    C(h)=d_{ABC}h^Ah^Bh^C=3h^1Q(h^I)+d_{111}(h^1)^3\ .
\end{equation}
No other cubic term can   be added, since it would be either of the same order in $h^1$, i.e.\ the classical term, or it would be forbidden by the quantum symmetry under which   $h^{I}\to -h^{I}$. For this argument to work, it is important that there is only an even number of vector multiplets, otherwise $-1_{n_V\times n_V}$ is not an element of SO($n_V-1,1$). In all our models $n_V$ turns out to be even. In fact, it was observed in \cite{Gkountoumis:2023fym} that this is  generic for freely acting orbifolds with $\mathcal{N}=2$ supersymmetry in $D=5$.

The next step is to argue that  
$d_{111}$ actually vanishes. This means that the result of integrating out all $A^1$-charged states in string theory is zero due to cancellations. The argument for the absence of quantum corrections uses again the dual pairs presented in section \ref{dual models with R-R fields}. The dual theories of the $n_V=4,6$ models are presented in the top two lines of \autoref{Table of dual pairs with R-R field}.
Duality enables us to determine the moduli space of one theory at strong coupling by going to weak coupling and large radius in a dual theory, as was explained in section \ref{dual pairs}. Classically, the vector multiplet moduli spaces for theories with $n_V>1$  are of the form \eqref{VMMod}. For theories without R-R scalars we have explicitly derived this result. For the dual theories, which do have R-R scalars, we have not shown this explicitly. However, as we discussed in section \ref{OSS}, the classical moduli space must be a subspace of $\text{E}_{6(6)}/\left[\text{Sp}(4)/\mathbb{Z}_2\right]$. This is again a homogeneous space and, combined with the constraints from $\mathcal{N}=2$ supergravity in $5D$ for $n_V=2,4,6$,  fixes classically the moduli space to be of the form \eqref{VMMod}\footnote{For higher (even) numbers of vector multiplets, there exist the special magic supergravities with moduli spaces which do not fit in \eqref{VMMod}. For example, the moduli spaces $\text{SL}(3,\mathbb{C})/\text{SU}(3)$ and $\text{SU}^*(6)/\text{USp}(6)$ can be obtained for $n_V=8$ and 14, respectively. These, however, always have massless R-R fields and are self-dual (in the sense of duality presented in section \ref{dual pairs}), but are worthwhile studying too.}.

Now, start with a theory $\mathcal{O}_{\text{IIB}}$ with no R-R scalars with classical vector multiplet moduli space $\mathcal{M}_V$, as given in \eqref{VMMod} and equipped with the canonical coset metric with the isometries inherited from $\mathcal{M}_V$ as a homogeneous space. The dual theory $\mathcal{O}_{\text{IIB}}'$ (with R-R scalars) has the same number of vector multiplets and by the arguments given above, it has the same classical vector multiplet moduli space and coset metric.  The same holds for the T-dual theory $\widetilde{\mathcal{O}}_{\text{IIA}}$. In order to examine if the moduli space metric on $\mathcal{M}_V$ is deformed by quantum corrections at strong coupling, in particular by the presence of a $d_{111}(h^1)^3$ term, we can simply go to the dual theory $\widetilde{\mathcal{O}}_{\text{IIA}}$ at weak coupling in $\widetilde{\lambda}_5$ and large radius $\tilde{\mathcal{R}}$ and study the moduli space there. But this is a regime where the classical answer can be trusted. Since the classical moduli space of $\widetilde{\mathcal{O}}_{\text{IIA}}$ is again homogeneous, with the same isometries as in the classical $\mathcal{O}_{\text{IIB}}$ theory, the coset metrics must be the same. But since one is the strong coupling version of the other, there cannot be any quantum corrections. 

Another way of saying this is that the real special geometry with the $d_{111}(h^1)^3$ term included does not fall into the classification of homogeneous spaces \cite{deWit:1991nm}, as one can check using (\ref{changex}). This would imply that the quantum moduli space would not be homogeneous.  But the dual theory describing the strong coupling regime has  in fact a homogeneous moduli space. Consequently, $d_{111}$ must be zero\footnote{In a theory with the $d_{111}(h^1)^3$ term included, at strong string coupling, the $d_{111}(h^1)^3$ term would dominate over the terms of order $h^1$. A theory with a cubic form $C(h)$ proportionally to $(h^1)^3$ only, would describe a theory with no vector multiplets, which would then be   supergravity coupled to hypermultiplets, and this is in contradiction with duality.}.

\section{Discussion}
\label{conclusions}

The main result of this paper was the determination of the moduli spaces arising in freely acting asymmetric orbifolds, and establishing  that in the cases we consider they are classically exact with no quantum corrections to the metric on the vector multiplet and hypermultiplet moduli space. For the vector multiplet moduli space metric, this led to a non-trivial prediction that no Chern-Simons terms in the Kaluza-Klein vector are induced by quantum corrections. It would be interesting to verify this by an explicit string theory calculation. Such a calculation could involve integrating out non-perturbative states, as the prediction relies on U-duality. In fact, one can verify that integrating out only low-mass charged states, as in \cite{Hull:2020byc}, does not lead to vanishing Chern-Simons terms in general.

Our models have an underlying $\mathcal{N}=8$ supersymmetry, which is spontaneously broken to $\mathcal{N}=2$ in the cases studied here. The theory
is described by a freely acting orbifold, which yields a spectrum of spontaneously broken $\mathcal{N}=8$ multiplets. The absence of quantum corrections in the $\mathcal{N}=2$ moduli space is partly due to this hidden supersymmetry, and fits in with the duality symmetries present in the $\mathcal{N}=8, D=5$ theories, so that the moduli spaces are subspaces of $\text{E}_{6(6)}/\left[\text{Sp}(4)/\mathbb{Z}_2\right]$. The arguments given in section \ref{EFFSUGRA}, however, assumed the absence of accidental massless modes, but some of the models we discussed in this paper do have such accidental massless modes. We gave more general arguments in sections \ref{dual pairs} and \ref{absence}, using comparison with dual theories. It would be very interesting to find out to what extent and under which assumptions our ideas can be applied more generally, for example to   four dimensional models and/or to models with completely broken supersymmetry.

Recently, there has been some work with similar mechanisms leading to the absence of quantum corrections \cite{Palti:2020qlc} in (special loci of) the moduli space in four-dimensional theories. While there are some differences with our work, the main claim of \cite{Palti:2020qlc}, that the absence of quantum corrections beyond those expected from the unbroken supersymmetry must be related to a higher supersymmetric theory, holds here too. In our case this is clear as the $\mathcal{N}=2$ freely acting orbifolds arise at special points in the moduli space of string theories with duality twists with $\mathcal{N}=8$ supersymmetry spontaneously broken to $\mathcal{N}=2$ and with duality symmetries that are subgroups of the U-duality group of maximally supersymmetric string theories. The precise connection to \cite{Palti:2020qlc}, and e.g.\ more recently \cite{Heckman:2024obe}, remains to be understood. For this, we would need to extend our work to four dimensions. This is clearly an interesting topic for future research.

\section*{Acknowledgements}

This work was initiated during a two-week visit at the Harvard Swampland Initiative. G.G. and S.V. are grateful for the warm hospitality, financial support and for stimulating discussions, in particular with Kaan Baykara and Cumrun Vafa. 
C.H. was supported by the STFC Consolidated Grants ST/T000791/1 and and ST/X000575/1, and
his research was supported in part by grant NSF PHY-2309135 to the Kavli Institute for Theoretical Physics (KITP).

\appendix
\label{appendix spectrum and table}
\section{Table of models and orbifold spectrum}

\subsection{Table of models}

\begin{table}[ht!]
\centering
{\renewcommand*{\arraystretch}{2.5}
\small{
	\begin{tabular}{|c|c|c|c|c|c|c|}
 \hline
 $\mathbb{Z}_p$	&  $(m_1,m_2,m_3)$ & $\tilde{u},u$ & $\Gamma^{4,4}(\mathcal{G})$ & $(n_V,n_H)$ & $\mathcal{M}_V$ & $\mathcal{M}_H$ \\ \hline
   $\mathbb{Z}_{12}$ & $(\tfrac{2\pi}{3},\tfrac{\pi}{2},\tfrac{\pi}{3})$ & $(\tfrac{1}{2},\tfrac{1}{6}),(\tfrac{1}{4},\tfrac{1}{4})$ & $\Gamma^{4,4}(D_4)$ & $(2,0)$ & $\mathbb{R}^+\times \mathbb{R}^+ $ & $-$ \\ \hline 
    $\mathbb{Z}_{12}$ & $(\tfrac{2\pi}{3},\tfrac{\pi}{2},\tfrac{2\pi}{3})$ & $(\tfrac{2}{3},0),(\tfrac{1}{4},\tfrac{1}{4})$ & $\Gamma^{4,4}(D_4)$ & $(4,0)$ & $\mathbb{R}^+\times \tfrac{\text{SO}(3,1)}{\text{SO}(3)} $ & $-$ \\ \hline 
   $\mathbb{Z}_6$ & $(\pi,\tfrac{2\pi}{3},\pi)$ & $(1,0),(\tfrac{1}{3},\tfrac{1}{3})$ & $\Gamma^{4,4}(A_2\oplus A_2)$ & $(6,0)$ & $\mathbb{R}^+\times \tfrac{\text{SO}(5,1)}{\text{SO}(5)} $ & $-$ \\ \hline 
 $\mathbb{Z}_6$ & $(\tfrac{\pi}{3},\tfrac{2\pi}{3},\tfrac{\pi}{3})$ & $(\tfrac{1}{3},0),(\tfrac{1}{3},\tfrac{1}{3})$ & $\Gamma^{4,4}(A_2\oplus A_2)$ & $(4,1)$ & $\mathbb{R}^+\times \tfrac{\text{SO}(3,1)}{\text{SO}(3)} $ & $\tfrac{\text{SU}(1,2)}{\text{U}(2)}$ \\ \hline 
		$\mathbb{Z}_6$ & $(\pi,\tfrac{\pi}{3},\tfrac{2\pi}{3})$ & $(\tfrac{5}{6},\tfrac{1}{6}),(\tfrac{1}{6},\tfrac{1}{6})$ & $\Gamma^{4,4}(A_2\oplus A_2)$ & $(2,2)$ & $\mathbb{R}^+\times \mathbb{R}^+$ & $\tfrac{\text{SU}(2,2)}{\text{SU}(2)\times \text{SU}(2)\times \text{U}(1)}$ \\ \hline 
  $\mathbb{Z}_4$ & $(\tfrac{\pi}{2},\pi,\tfrac{\pi}{2})$ & $(\tfrac{1}{2},0),(\tfrac{1}{2},\tfrac{1}{2})$ & $\Gamma^{4,4}(D_2\oplus D_2)$ & $(4,2)$ & $\mathbb{R}^+\times \tfrac{\text{SO}(3,1)}{\text{SO}(3)} $ & $\tfrac{\text{SU}(2,2)}{\text{SU}(2)\times \text{SU}(2)\times \text{U}(1)}$ \\ \hline 
       	\end{tabular}}}
        \caption{\textit{Table of $\mathcal{N}=2$ models in $5D$ with no R-R massless states. In all models $m_4=0$. These models are discussed in detail in section \ref{N=2 models in 5d with exact moduli spaces}. Also, recall that $D_2\cong A_1\oplus A_1$.}}
        \label{Table of models}
\end{table}

\newpage
\subsection{Orbifold spectrum}
\label{appendix orbifold spectrum}

\renewcommand{\arraystretch}{1.4}
\begin{table}[ht!]
\centering
\begin{tabular}{|c|c|c|c|c|}
\hline
\;Sector\; & State & Orbifold charge & SO$(3)$ rep & SO$(4)$ rep \\ \hline\hline
NS-NS & \;\;$\tilde{b}^{\hat{\mu}}_{-1/2}\ket{0}_L \otimes {b}^{\hat{\nu}}_{-1/2}\ket{0}_R$\;\; & $1$ & $\textbf{5} \oplus 3\times\textbf{3}\oplus2\times \textbf{1}$ & $(\textbf{3}\oplus\textbf{1},\textbf{3}\oplus\textbf{1})$ \\ \cline{2-5}
& $\tilde{b}^{\hat{\mu}}_{-1/2}\ket{0}_L \otimes {b}^i_{-1/2}\ket{0}_R$ & $e^{i(m_2\pm m_4)}$ & {$2\times{\textbf{3}\oplus2\times\textbf{1}}$} & $2\times({\textbf{2}},{\textbf{2}})$ \\ \cline{2-5}
& $\tilde{b}^{\hat{\mu}}_{-1/2}\ket{0}_L \otimes \bar{{b}}^i_{-1/2}\ket{0}_R$ & $e^{-i(m_2\pm m_4)}$ & {$2\times{\textbf{3}\oplus2\times\textbf{1}}$} & $2\times({\textbf{2}},{\textbf{2}})$ \\ \cline{2-5}
& $\tilde{b}^{i}_{-1/2}\ket{0}_L \otimes {b}^{\hat{\mu}}_{-1/2}\ket{0}_R$ & $e^{i(m_1\pm m_3)}$ & {$2\times{\textbf{3}\oplus2\times\textbf{1}}$} & $2\times({\textbf{2}},{\textbf{2}})$ \\ \cline{2-5}
& $\bar{\tilde{b}}^{i}_{-1/2}\ket{0}_L \otimes {b}^{\hat{\mu}}_{-1/2}\ket{0}_R$ & $e^{-i(m_1\pm m_3)}$ & {$2\times{\textbf{3}\oplus2\times\textbf{1}}$} & $2\times({\textbf{2}},{\textbf{2}})$ \\ \cline{2-5}
& $\tilde{b}^{i}_{-1/2}\ket{0}_L \otimes {b}^j_{-1/2}\ket{0}_R$ & $e^{i(m_1\pm m_3)+i(m_2\pm m_4)}$ & {$4\times\textbf{1}$} & $4\times({\textbf{1}},{\textbf{1}})$ \\ \cline{2-5}
& $\tilde{b}^{i}_{-1/2}\ket{0}_L \otimes \bar{{b}}^j_{-1/2}\ket{0}_R$ & $e^{i(m_1\pm m_3)-i(m_2\pm m_4)}$ & {$4\times\textbf{1}$} & $4\times({\textbf{1}},{\textbf{1}})$ \\ \cline{2-5}
& $\bar{\tilde{b}}^{i}_{-1/2}\ket{0}_L \otimes {b}^j_{-1/2}\ket{0}_R$ & $e^{-i(m_1\pm m_3)+i(m_2\pm m_4)}$ & {$4\times\textbf{1}$} & $4\times({\textbf{1}},{\textbf{1}})$ \\ \cline{2-5}
& $\bar{\tilde{b}}^{i}_{-1/2}\ket{0}_L \otimes \bar{{b}}^j_{-1/2}\ket{0}_R$ & $e^{-i(m_1\pm m_3)-i(m_2\pm m_4)}$ & {$4\times\textbf{1}$} & $4\times({\textbf{1}},{\textbf{1}})$ \\ \hline\hline

R-R & $|a_{1,2}\rangle_L \otimes |a_{1,2}\rangle_R$ & $e^{\pm im_1\pm im_2}$ & {$4\times{\textbf{3}\oplus4\times\textbf{1}}$} & $4\times({\textbf{3}\oplus\textbf{1}},{\textbf{1}})$ \\ \cline{2-5}
& $|a_{1,2}\rangle_L \otimes |a_{3,4}\rangle_R$ & $e^{\pm im_1\pm im_4}$ & {$4\times{\textbf{3}\oplus4\times\textbf{1}}$} & $4\times({\textbf{2}},{\textbf{2}})$ \\ \cline{2-5}
& $|a_{3,4}\rangle_L \otimes |a_{1,2}\rangle_R$ & $e^{\pm im_3\pm im_2}$ & {$4\times{\textbf{3}\oplus4\times\textbf{1}}$} & $4\times({\textbf{2}},{\textbf{2}})$ \\ \cline{2-5}
& $|a_{3,4}\rangle_L \otimes |a_{3,4}\rangle_R$ & $e^{\pm im_3\pm im_4}$ & {$4\times{\textbf{3}\oplus4\times\textbf{1}}$} & $4\times({{\textbf{1},\textbf{3}\oplus\textbf{1}}})$ \\ \hline\hline

NS-R & $\tilde{b}^{\hat{\mu}}_{-1/2}\ket{0}_L \otimes |a_{1,2}\rangle_R$ & $e^{\pm im_2}$ & {$2\times{\textbf{4}\oplus4\times\textbf{2}}$} & $2\times({\textbf{3}\oplus\textbf{1}},{\textbf{2}})$ \\ \cline{2-5}
& $\tilde{b}^{\hat{\mu}}_{-1/2}\ket{0}_L \otimes |a_{3,4}\rangle_R$ & $e^{\pm im_4}$ & {$2\times{\textbf{4}\oplus4\times\textbf{2}}$} & $2\times({\textbf{2}},{\textbf{3}\oplus\textbf{1}})$ \\ \cline{2-5}
& $\tilde{b}^{i}_{-1/2}\ket{0}_L \otimes |a_{1,2}\rangle_R$ & $e^{i(m_1\pm m_3)\pm im_2}$ & {$4\times\textbf{2}$} & $4\times({\textbf{2}},{\textbf{1}})$ \\ \cline{2-5}
& $\tilde{b}^{i}_{-1/2}\ket{0}_L \otimes |a_{3,4}\rangle_R$ & $e^{i(m_1\pm m_3)\pm im_4}$ & {$4\times\textbf{2}$} & $4\times({\textbf{1}},{\textbf{2}})$ \\ \cline{2-5}
& $\bar{\tilde{b}}^{i}_{-1/2}\ket{0}_L \otimes |a_{1,2}\rangle_R$ & $e^{-i(m_1\pm m_3)\pm im_2}$ & $4\times\textbf{2}$ & $4\times({\textbf{2}},{\textbf{1}})$ \\ \cline{2-5}
& $\bar{\tilde{b}}^{i}_{-1/2}\ket{0}_L \otimes |a_{3,4}\rangle_R$ & $e^{-i(m_1\pm m_3)\pm im_4}$ & {$4\times\textbf{2}$} & $4\times({\textbf{1}},{\textbf{2}})$ \\ \hline\hline

R-NS & $|a_{1,2}\rangle_L \otimes {b}^{\hat{\mu}}_{-1/2}\ket{0}_R$ & $e^{\pm im_1}$ & {$2\times{\textbf{4}\oplus4\times\textbf{2}}$} & $2\times({\textbf{3}\oplus\textbf{1}},{\textbf{2}})$ \\ \cline{2-5}
& $|a_{3,4}\rangle_L \otimes {b}^{\hat{\mu}}_{-1/2}\ket{0}_R$ & $e^{\pm im_3}$ & {$2\times{\textbf{4}\oplus4\times\textbf{2}}$} & $2\times({\textbf{2}},{\textbf{3}\oplus\textbf{1}})$ \\ \cline{2-5}
& $|a_{1,2}\rangle_L \otimes {b}^i_{-1/2}\ket{0}_R$ & $e^{\pm im_1+i(m_2\pm m_4)}$ & {$4\times\textbf{2}$} & $4\times({\textbf{2}},{\textbf{1}})$ \\ \cline{2-5}
& $|a_{3,4}\rangle_L \otimes {b}^i_{-1/2}\ket{0}_R$ & $e^{\pm im_3+i(m_2\pm m_4)}$ & {$4\times\textbf{2}$} & $4\times({\textbf{1}},{\textbf{2}})$ \\ \cline{2-5}
& $|a_{1,2}\rangle_L \otimes \bar{{b}}^i_{-1/2}\ket{0}_R$ & $e^{\pm im_1-i(m_2\pm m_4)}$ & {$4\times\textbf{2}$} & $4\times({\textbf{2}},{\textbf{1}})$ \\ \cline{2-5}
& $|a_{3,4}\rangle_L \otimes \bar{{b}}^i_{-1/2}\ket{0}_R$ & $e^{\pm im_3-i(m_2\pm m_4)}$ & {$4\times\textbf{2}$} & $4\times({\textbf{1}},{\textbf{2}})$ \\ \hline
\end{tabular}
\captionsetup{width=.9\linewidth}
\caption{\textit{The spectrum of lowest excited string states in the untwisted orbifold sector, including their orbifold charge and representations under the massless little group} SO$(3)\approx\text{SU}(2)$ \textit{and the massive little group} SO$(4)\approx\text{SU}(2)\times\text{SU}(2)$ \textit{in five dimensions. We use the indices $\hat{\mu},\hat{\nu}=0,\ldots,5$ for the real coordinates on $\mathbb{R}^{1,4}\times S^1$ and the indices $i,j=1,2$ for the complex coordinates on $T^4$. This table is taken from \cite{Gkountoumis:2023fym}.}}
\label{huiberttable}
\end{table}
\renewcommand{\arraystretch}{1}

\bibliographystyle{unsrt}
\bibliography{bib}

\end{document}